%% file: strat.tex
\documentclass[11pt]{article}
\usepackage{epsfig,sint,macros,cite,mathbbol}
\input axodraw.sty
\begin{document}
\input title.tex
\input sect1.tex

\input sect2.tex

\input sect3.tex

\input sect4.tex

\input sect5.tex

\input sect6.tex

\input sect7.tex

\input sect8.tex
\newpage
\begin{appendix}
\input appendix1.tex

\input appendix2.tex

\input appendix3.tex

\input appendix4.tex
\end{appendix}

\input biblio.tex
\end{document}

%% file: title.tex
\begin{titlepage}
\begin{flushright}
\begin{minipage}[t]{6cm}
BI-TP 2004/14    \hfill FTUV-04-0705     \\
CPT-2004/P.031   \hfill IFIC/04-29       \\
DESY 04-081      \hfill MPP-2004-55
\end{minipage}
\end{flushright}

\vskip 0.5 cm
\begin{center}
  {\Large\bf 
  A strategy to study the r\^ole of the charm quark\\
  in explaining the $\Delta{I}=1/2$ rule \\[0.5ex] } 
\end{center}
\vskip 0.5 cm
\begin{center}
{\large 
     L.~Giusti$^{\scriptscriptstyle a}$,
     P.~Hern\'andez$^{\scriptscriptstyle b}$,
     M.~Laine$^{\scriptscriptstyle c}$,
     P.~Weisz$^{\scriptscriptstyle d}$,
     H.~Wittig$^{\scriptscriptstyle e}$
}
\vskip 0.5cm
$^{\scriptstyle a}$
Centre de Physique Th\'eorique, Case 907, CNRS Luminy\\
F-13288 Marseille Cedex 9, France
\vskip 1.5ex
$^{\scriptstyle b}$
Dpto. F\'isica Te\'orica and IFIC, Edificio Institutos Investigaci\'on, \\
Apt. 22085, E-46071 Valencia, Spain
\vskip 1.5ex
$^{\scriptstyle c}$
Faculty of Physics, University of Bielefeld, \\
D-33501 Bielefeld, Germany
\vskip 1.5ex
$^{\scriptstyle d}$
Max-Planck-Institut f\"ur Physik, F\"ohringer Ring 6,\\
D-80805 Munich, Germany
\vskip 1.5ex
$^{\scriptstyle e}$
DESY, Theory Group, Notkestrasse 85,\\
D-22603 Hamburg, Germany
\vskip 1.0cm
{\bf Abstract}
\vskip 0.35ex
\end{center}

\noindent
We present a strategy designed to separate several possible origins of
the well-known enhancement of the $\Delta{I}=1/2$ amplitude in
non-leptonic kaon decays. In particular, we seek to unambiguously
quantify the r\^ole of the charm quark mass in the observed
enhancement. This is achieved by considering QCD with an unphysically
light charm quark, so that the theory possesses an approximate
$\fourby$ chiral symmetry. The strategy proceeds by computing the
relevant operator matrix elements and monitoring their values as the
charm quark mass departs from the SU(4)-symmetric situation. We study
the influence of the charm quark mass in Chiral Perturbation Theory.
First results from lattice simulations in the SU(4)-symmetric limit
are also discussed.

\vfill

\begin{center}
December 2004
\end{center}

\eject

\vfill

\eject

\end{titlepage}

%% file: sect1.tex
\section{Introduction}

A quantitative understanding of non-leptonic kaon decays, such as
$K\to\pi\pi$, has been elusive for many years, and thus the
explanation of the famous $\Delta{I}=1/2$ rule or the value of
$\epsilon^\prime/\epsilon$ has remained a longstanding problem. Early
analyses have shown that, in a Standard Model-based explanation, the
bulk of the enhancement of the $\Delta{I}=1/2$ amplitude must be due
to long-distance contributions generated by the strong interactions
\cite{gammapm0}. A reliable determination of these effects must
inevitably be performed at the non-perturbative level
\cite{WME84,CPS_symm}. Currently, lattice simulations of QCD are the
only known methodology which can achieve this goal with controlled
systematic errors. Lattice studies of $K\to\pi\pi$ decays have,
however, been hampered by a number of technical difficulties, for
instance by the so-called Maiani-Testa No-Go Theorem \cite{MaiTes90},
which states that transition amplitudes of two-body decays cannot
simply be obtained from the asymptotic behaviour of Euclidean
correlation functions computed in very large volumes. In addition,
lattice calculations employing commonly used Wilson fermions need
non-perturbative subtractions of power divergences for defining
properly renormalised operators which enter the effective weak
Hamiltonian \cite{powsub}.\footnote{For recent progress using twisted
mass QCD, see \cite{Kpipi_tmQCD} and \cite{Kpipi_tmQCD2}.} Thus,
efforts to investigate the $\Delta{I}=1/2$ rule numerically on the
lattice had practically come to a halt during most part of the 1990s.

Recently, it was realised, though, that the Maiani-Testa Theorem does
not apply to volumes with linear extensions of a few\,\fm, and a
relation between the physical two-body decay rate and the square of
the transition matrix element in finite volume could be derived
\cite{LelLuesch00} (for subsequent work, see \cite{RoSo02}).
Furthermore, the advent of fermionic discretisations which preserve
chiral symmetry at non-zero lattice spacing (Ginsparg-Wilson fermions)
\cite{GinsWil,Kaplan,Shamir,FurSha,Hasenfratz,HLN,NeubergerDirac,ExactChSy,Locality},
has alleviated the mixing problem to the extent that the
renormalisation patterns of four-fermion operators mediating
$\Delta{S}=1$ transitions are like in the
continuum\,\cite{SteLeo01}. These developments have removed the main
obstacles for a lattice treatment of non-leptonic kaon decays, based
on first principles. An approximate realisation of Ginsparg-Wilson
fermions via the so-called Domain Wall formulation has already been
applied to non-leptonic kaon decays with some success
\cite{RBC_kpipi,CPPACS_kpipi}.

The decay of a neutral kaon into a pair of pions in a state with
isospin~$I$ is described by the transition amplitude
\be
   T(K^0\to\pi\pi\big|_{I=\alpha}) = i A_\alpha\rme^{i\delta_\alpha},
   \qquad\alpha=0,\,2,
\ee
where $\delta_\alpha$ is the scattering phase shift. In this paper 
we focus on the $\Delta{I}=1/2$ rule, i.e. the empirical observation 
that the amplitude $|A_0|$ is significantly larger than $|A_2|$,
\be
   |A_0|/|A_2| = 22.1.
\ee
There are several possible sources for an enhancement due to strong
interaction effects. These involve physics at the scale of the charm
quark, i.e. at around 1\,\GeV, physics at an intrinsic QCD scale
$E_{\rm QCD}\approx250\,\MeV$, and effects due to pionic final state
interactions at around 100\,\MeV. It remains unclear, though, whether
the enhancement is the result of an accumulation of several effects,
each giving a moderate contribution, or whether it is mainly due to a
single cause or mechanism.

The possible r\^ole of the charm quark has been pointed out long ago
in ref. \cite{SVZ77}: for energy scales $E\ll m_c$ the GIM mechanism
no longer operates, which gives rise to so-called ``penguin
operators'' that mediate $\Delta{I}=1/2$ transitions. The commonly
accepted scenario that the $\Delta{I}=1/2$ rule arises predominantly
from long-distance QCD contributions has been supported by several
analytical \cite{BarBuGer87,KamMissWyl90,NeuSte91,Pich90,Bertoetal}
and numerical \cite{RBC_kpipi,CPPACS_kpipi} studies.

In this paper we describe a general strategy which allows to
disentangle contributions from the various sources and quantify them
using numerical simulations. The main idea is to keep an active charm
and determine the leading low-energy constants (LECs) associated with
the CP conserving $\Delta S=1$ weak Hamiltonian of the chiral
low-energy effective theory as a function of the charm quark
mass. These parameters can be determined by computing suitable
correlation functions at small masses and momenta in a lattice
simulation and comparing them with the predictions of Chiral
Perturbation Theory (ChPT). If the charm is degenerate with the light
quarks, $m_u=m_d=m_s=m_c$, the theory has an exact $\rm
SU(4)_{L}\times SU(4)_{R}$ symmetry in the chiral limit, which is
broken explicitly by the weak interactions. In this situation the
calculation of the LECs corresponding to $\Delta{I}=1/2$ and 3/2
transitions will expose any QCD contribution to the enhancement at
scales around $E_{\rm{QCD}}$. The effect of a heavier charm quark can
then be isolated by monitoring the amplitudes as $m_c$ departs from
the degenerate limit: $m_c> m_u=m_d=m_s$. By keeping an active charm
quark and employing a lattice formulation which preserves chiral
symmetry, power-divergent subtractions of the relevant operators can
be avoided, and small quark masses can be simulated without numerical
instabilities.

Since simulations with dynamical Ginsparg-Wilson fermions are still
prohibitively expensive, it is reasonable to perform initial tests of
our proposed strategy in the quenched approximation. It is well known,
though, that the quenched theory is afflicted with several problems,
whose implications are discussed in detail in the relevant sections
below. In order to illustrate our strategy, we present first numerical
results for the SU(4)-symmetric case obtained in the quenched
approximation for quark masses near $m_s/2$. Numerical results for
smaller masses and the investigation of the dependence on the charm
quark mass are left to future publications.

The outline of the remainder of this paper is as follows: in
Sect.\,\ref{sec_Hw_cont} we discuss the effective weak interactions
with an active charm quark. The lattice-regularised theory, formulated
using Ginsparg-Wilson quarks, is described in
Sect.~\ref{sec_Hw_latt}. The renormalisation group invariant
formulation of the effective weak Hamiltonian is addressed in
Sect.~\ref{sec_RGI}. In Sect.~\ref{sec_ChPT} we discuss the effective
low-energy description of $\Delta{S}=1$ weak interactions in terms of
ChPT in a finite volume. The decoupling of the charm quark for small
quark masses, i.e. $m_c\;\gtaeq\;m_u, m_d, m_s$, is analysed in
Sect.~\ref{sec_charm_ChPT} using ChPT. Our numerical results are
discussed in Sect.~\ref{sec_num}, and in Sect.~\ref{sec_concl} we
present our conclusions. In several appendices we provide further
information on the SU(4) classification of operators, the
transformation properties of four-quark operators in the lattice
theory, the perturbative renormalisation of four-quark operators using
overlap fermions, as well as details of the calculations performed in
finite-volume ChPT.

%% file: sect2.tex
\section{The effective weak interactions with an active charm quark
\label{sec_Hw_cont}}

In this section we collect some basic facts and definitions regarding
the $\Delta{S}=1$ effective weak interaction, focusing on the less
familiar case of an active charm quark. Throughout this paper we work
in Euclidean space-time.

\subsection{Operator product expansion and global symmetries}

The decay of a (neutral or charged) kaon into two pions is induced by
charged-current weak interactions, mediated via the exchange of a
$W$-boson. It can be described in terms of an effective $V-A$
current-current interaction, i.e.
\be
   S_{\rm w}=
   {\textstyle\frac{1}{2}}g_{\rm w}^2\sum_{q=u,c,t}(V_{qs})^*V_{qd}
   \int\rmd^4x\,\rmd^4y\,(\sbar\gamma_\mu P_{-}q)(x)\,D_{\mu\nu}(x-y)\, 
                         (\qbar\gamma_\nu P_{-}d)(y),
\label{eq_Sw}
\ee
where $g_{\rm{w}}^2=4\sqrt{2}G_{\rm F} M_W^2$, $V_{qs},\,V_{qd}$
denote elements of the CKM matrix, $P_{-}=\frac{1}{2}(1-\gamma_5)$,
and $D_{\mu\nu}$ is the propagator of the $W$-boson. Contributions
from the top-quark are suppressed by three orders of magnitude
relative to those from the up-quark and can thus be safely neglected.
At this level of accuracy one has
$(V_{us})^*V_{ud}=-(V_{cs})^*V_{cd}$, so that
\bea
 S_{\rm w}&=&{\textstyle\frac{1}{2}}g^2_{\rm{w}}(V_{us})^*V_{ud}
    \int\rmd^4x\,\rmd^4y \nonumber\\
   & &\times\Big\{(\sbar\gamma_\mu P_{-}u)(x)\,D_{\mu\nu}(x-y)\, 
                  (\ubar\gamma_\nu P_{-}d)(y) -(u\to c)\Big\}.
\eea
The dominant contribution to $S_{\rm w}$ comes from the region
$x\approx{y}$, so that the integral can be evaluated using the
operator product expansion:
\be
  S_{\rm w}\approx\int\rmd^4x\,{\cal H}_{\rm w}(x),\qquad
  {\cal H}_{\rm w}(x)=\frac{g^2_{\rm{w}}}{4M_W^2}(V_{us})^*V_{ud}
  \sum_n k_n{\cal Q}_n(x),
\label{eq_Sw_OPE}
\ee
where the coefficients $k_n$ depend on the $W$-boson mass, $M_W$, and
the renormalisation scheme used to define ${\cal Q}_n$. In order to
classify the operators that can occur in the sum, we now discuss the
global symmetries that must be respected.

To this end we consider QCD for two generations and write its action
as
\be
   S=S_{\rm G} + \int\rmd^4x\,\psibar(x)
                 \left(D+MP_{+}+M^\dagger P_{-}\right)\psi(x),
\ee
where $S_{\rm G}$ denotes the gauge action, $D$ the massless Dirac
operator, $M$ the quark mass matrix, and $\psi$ the four-flavour quark
field, with flavour components $u,d,s,c$. The action is invariant
under $\rm SU(4)_L\times SU(4)_R$ chiral transformations if $M$ is
transformed according to the $\bf 4\otimes4^*$ representation. For
real diagonal $M$ it is also invariant under parity and charge
conjugation.

Under $\rm SU(4)_L\times SU(4)_R$ the effective action $S_{\rm w}$
divides into two parts that belong to different representations. The
decomposition reads
\bea
 & & S_{\rm w}=S^{+}_{\rm w}+S^{-}_{\rm w},  \nonumber \\
 & & S^{\pm}_{\rm w}={\textstyle\frac{1}{4}}g^2_{\rm{w}}(V_{us})^*V_{ud}
     \int\rmd^4x\,\rmd^4y\Big\{
     [{\cal P}_{suud}(x,y)\pm{\cal P}_{sudu}(x,y)]-(u\to c)\Big\},
\eea
where
\be
   {\cal P}_{\alpha\beta\gamma\delta}(x,y) =
   \big(\psibar_\alpha\gamma_\mu P_{-}\psi_\gamma\big)(x)\,
   D_{\mu\nu}(x-y)\,
   \big(\psibar_\beta\gamma_\nu P_{-}\psi_\delta\big)(y),
\ee
with generic flavour indices $\alpha,\ldots,\delta$. From the
structure of ${\cal P}_{\alpha\beta\gamma\delta}$ it is evident that
both parts are singlets under $\rm SU(4)_R$, while it can be shown
that $S_{\rm w}^{+}$ and $S_{\rm w}^{-}$ transform according to the
irreducible representations of $\rm SU(4)_L$ of dimensions~84 and~20
respectively \cite{georgi}. Furthermore, the effective action, as well
as the QCD action, are invariant under the CPS transformation,
i.e. combined operations of CP followed by an interchange of $d$ and
$s$ quarks \cite{CPS_symm}.

Having thus listed the global symmetries, we now proceed to find the
operators of dimension $d\le6$ which occur in the operator product
expansion, \eq{eq_Sw_OPE}, noting that operators of higher dimension
are suppressed by powers of $M_W$. We start the discussion with
four-quark operators; given a suitable basis of operators, we seek to
construct linear combinations that are $\rm SU(4)_R$ singlets and
which transform according to irreducible representations of $\rm
SU(4)_L$. The result is
\bea
& & {\cal Q}_1^{\pm} = ([{O}_1]_{suud}\pm[{O}_1]_{sudu})
                      -(u\to c),
    \label{eq_Q1pm} \\
& & [{O}_1]_{\alpha\beta\gamma\delta}\equiv
    \big(\psibar_\alpha\gamma_\mu P_{-}\psi_\gamma\big)
    \big(\psibar_\beta\gamma_\mu  P_{-}\psi_\delta\big).
    \label{eq_O1}
\eea
Note that the decomposition yields unique operators
${\cal{Q}}_1^{+},\,{\cal Q}_1^{-}$, which transform according to
irreducible representations of dimensions 84 and 20,
respectively. Further details are described in
Appendix~\ref{app1}. Furthermore, the requirement of CPS symmetry is
also fulfilled.

Two-quark operators of dimension $d\le6$ which are singlets under
$\rm SU(4)_R$ can be constructed by including derivatives and two or
more factors of the mass matrix. Dropping operators that are related
via the equations of motion and operators transforming according to
low-dimensional representations of $\rm SU(4)_L$ one is left with
\cite{notes:mixing}
\be
   \big(MM^\dagger\big)_{\gamma\beta}
   \big(\psibar_\alpha P_{+}[M\psi]_\delta\big)
   \quad\hbox{and}\quad
   \big(MM^\dagger\big)_{\gamma\beta}
   \big([\psibar M^\dagger]_\alpha P_{-}\psi_\delta\big).
\ee
The projected operators that transform according to representations
of dimension 84 and 20 and which satisfy CPS symmetry are then
\bea
& & {\cal Q}_2^{\pm} = ([{O}_2]_{suud}\pm[{O}_2]_{sudu})
                      -(u\to c), \label{eq_Q2pm} \\
& & [{O}_2]_{\alpha\beta\gamma\delta}\equiv
   \big(MM^\dagger\big)_{\gamma\beta}\big\{
   \big(\psibar_\alpha P_{+}[M\psi]_\delta\big)+
   \big([\psibar M^\dagger]_\alpha P_{-}\psi_\delta\big)
   \big\}
 \nonumber\\
& & \phantom{[{O}_2]_{\alpha\beta\gamma\delta}\equiv\,}
    +\big(\alpha\leftrightarrow\beta,\,\gamma\leftrightarrow\delta\big).
\eea
For $M={\rm diag}(m_u,m_d,m_s,m_c)$ the expression for ${\cal Q}_2^{\pm}$
is
\be
   {\cal Q}_2^{\pm}=(m_u^2-m_c^2)\big\{
    m_d(\sbar P_{+}d)+m_s(\sbar P_{-}d) \big\}.
\label{eq_Q2def}
\ee
Zero-quark operators which transform under $\rm SU(4)_L$ can also be
constructed using at least four powers of the mass matrix. Any such
terms conserve flavour automatically (the mass matrix is diagonal) and
hence they do not contribute to the part of the effective weak
Hamiltonian we are interested in.

\subsection{Renormalisation and mixing}

So far we have neglected the issue of renormalisation. We note that
the effective action $S_{\rm w}$ in \eq{eq_Sw} is finite and does not
need any subtractions. This is obvious from the fact that the charged
currents are not anomalous in QCD and thus have a natural
normalisation. After the operator product expansion, the operators
${\cal Q}_1^{\pm}$ and ${\cal Q}_2^{\pm}$ of dimension 6 that appear
in \eq{eq_Sw_OPE} must be renormalised, and there is no reason why
operators with the same transformation properties should not mix.
Assuming that the adopted regularisation preserves enough of the
relevant symmetry structure to exclude any other mixings, the
relations between renormalised and bare operators are of the general
form
\bea
 & & {\cal Q}_1^{\pm} =
     {\cal Z}_{11}^{\pm}{\cal Q}_1^{\pm,\rm{bare}}
    +{\cal Z}_{12}^{\pm}{\cal Q}_2^{\pm,\rm{bare}}, \nonumber\\
 & & {\cal Q}_2^{\pm} =
     {\cal Z}_{21}^{\pm}{\cal Q}_1^{\pm,\rm{bare}}
    +{\cal Z}_{22}^{\pm}{\cal Q}_2^{\pm,\rm{bare}}.
\eea
The lattice formulation with Ginsparg-Wilson fermions discussed in the
next section is an example of a regularisation where these relations
hold without modification.

Since the operators ${\cal Q}_2^{\pm,\rm{bare}}$, defined as in
\eq{eq_Q2def}, are linear combinations of the non-singlet chiral
densities, which are multiplicatively renormalisable, we are free to
set ${\cal Z}_{21}^{\pm}=0$. The renormalised operators ${\cal
Q}_2^{\pm}$ are then obtained as linear combinations of renormalised
densities with coefficients that are polynomials in the renormalised
quark masses. As is obvious from eq. (\ref{eq_Q2def}), the
Glashow-Iliopoulos-Maiani (GIM) mechanism ensures that any
contribution proportional to ${\cal Q}_2^{\pm}$ vanishes when
$m_c=m_u$. In particular, the mixing of ${\cal Q}_2^{\pm}$ with ${\cal
Q}_1^{\pm}$ is absent in the chiral limit, and hence the factors
${\cal Z}_{11}^{\pm}$ can be fixed at vanishing quark masses. The
constants ${\cal Z}_{12}^{\pm}$ must then be determined by requiring
that any residual divergences in the matrix elements of ${\cal
Q}_1^{\pm}$ for $m_u\;{\neq}\;m_c$ are cancelled.

We note that the mixing of ${\cal Q}_1^{\pm}$ with ${\cal Q}_2^{\pm}$
is usually ignored in matrix elements in which there is no momentum
transfer between initial and final states, $|i\rangle$ and
$|f\rangle$. The reason for this is that the operators
$({\sbar}P_{\pm}d)$ can be written as a total four-divergence, and
hence the matrix element
\bea
& & \langle{f}|({\sbar}P_{\pm}d)|i\rangle =
    \nonumber\\ 
& & (m_s-m_d)^{-1}\,\partial_\mu\langle{f}|({\sbar}{\gamma_\mu}d)|i\rangle
 \pm(m_s+m_d)^{-1}\,\partial_\mu
    \langle{f}|({\sbar}{\gamma_\mu\gamma_5}d)|i\rangle
\eea
vanishes if $|i\rangle$ and $|f\rangle$ have the same
four-momentum. In other words, only the terms proportional to ${\cal
Q}_1^\pm$ contribute in physical matrix elements. Evidently this
argumentation is correct only if $m_s\pm m_d \neq 0$ and if kinematical
singularities at zero momentum transfer can be excluded.

%% file: sect3.tex
\section{The effective weak Hamiltonian in lattice QCD
  \label{sec_Hw_latt}} 

The construction of the effective weak Hamiltonian on the lattice
proceeds by finding a linear combination of local composite fields
with coefficients such that the correct operator (including
normalisation) is obtained in the continuum limit. If a lattice
formulation is used which preserves chiral symmetry, many of the
difficulties encountered with Wilson fermions, such as the mixing with
lower dimensional operators, can be avoided \cite{SteLeo01}. In
particular, one may require that the operator basis transforms in a
simple way under the chiral $\rm SU(4)_L\times SU(4)_R$ symmetry group
at non-zero lattice spacing.

\subsection{Ginsparg-Wilson fermions \label{sec_GWferms}}

The formulation of lattice QCD with exact chiral symmetry proceeds by
introducing a lattice Dirac operator $D$ which satisfies the
Ginsparg-Wilson relation
\be
   \gamma_5 D+D\gamma_5={\abar}D\gamma_5 D.
\ee
Explicitly we take $D$ to be the Neuberger-Dirac operator
\cite{NeubergerDirac}, defined by
\be
   D=\frac{1}{\abar}\left\{1-A(A^\dagger A)^{-1/2}\right\}, \qquad 
   A\equiv1+s-aD_{\rm w},
\label{eq_Dneu}
\ee
where $D_{\rm w}$ denotes the massless Wilson-Dirac operator, $a$ is
the lattice spacing, and $s$ is a free parameter in the range $|s|<1$,
which can be tuned to optimise the locality properties of $D$
\cite{Locality}. If $\abar$ is set to
\be
    \abar=\frac{a}{1+s}
\ee
it is straightforward to check that $D$ satisfies the Ginsparg-Wilson
relation. The infinitesimal chiral transformations of the quark fields
$\psi$ and $\psibar$ are given by \cite{ExactChSy}
\be
   \delta\psi=\lambda^a\gamma_5(1-{\abar}D)\psi,\qquad
   \delta\psibar=\psibar\gamma_5\lambda^a,
\label{eq_InfChirTrans}
\ee
where $\lambda^a$ is a flavour matrix. Furthermore, it is useful to
define the modified fermion field $\psitilde$ by
\be
   \psitilde=(1-{\textstyle\frac{1}{2}}{\abar}D)\psi,
\label{eq_psitilde}
\ee
whose infinitesimal chiral transformation is given by
\be
   \delta\psitilde=\lambda^a\gamma_5\psitilde.
\ee
An important reason for introducing the modified field $\psitilde$ is
that local composite operators constructed from $\psitilde$ and
$\psibar$ have simple transformation properties under the chiral
symmetry, similar to those in the continuum.

\subsection{Operator basis}

In the construction of the weak Hamiltonian in the lattice theory we
are concerned with finding a basis of local operators of dimension
$d\leq6$ with the same transformation behaviour as ${\cal Q}_1$ and
${\cal Q}_2$ under the exact symmetries of the lattice theory. These
include the discrete lattice symmetries, the gauge transformations,
$\fourby$ chiral transformations, and the CPS symmetry.

Concentrating first on four-quark operators, we note that in the
continuum the problem is solved by constructing a basis of
gauge-invariant operators that transform as scalar fields under the
(restricted) Lorentz group $\rm SO(4)$. For Wilson-type fermions the
classification proceeds along the same lines, except that the Lorentz
group $\rm SO(4)$ is reduced to the hypercubic group
$\hbox{SO}(4,\mathbb{Z})$. More precisely, the task is to find all
tensors that are invariant under the spin covering
$\hbox{Spin}(4,\mathbb{Z})$. A detailed analysis \cite{notes:4quark}
then shows that no additional invariant tensors can occur under
$\hbox{Spin}(4,\mathbb{Z})$, and hence the basis of four-quark
operators in the continuum theory is also a basis in the lattice
theory.

It is easy to convince oneself that four-quark operators on the
lattice composed in terms of $\psibar$ and the modified fields
$\psitilde$ have exactly the same transformation behaviour under
$\fourby$ as the corresponding operators in the continuum. Thus, the
lattice counterpart of the operator ${\cal Q}_1^\pm$ in
eq. (\ref{eq_Q1pm}) can be chosen as
\be
   {\cal Q}_1^{\pm,\rm bare} = \Big\{
   (\sbar\gamma_{\mu}P_{-}\tilde{u})(\ubar\gamma_{\mu}P_{-}\tilde{d})
\pm(\sbar\gamma_{\mu}P_{-}\tilde{d})(\ubar\gamma_{\mu}P_{-}\tilde{u})
   \Big\} - (u\,{\to}\,c).
\label{eq_Q1_bare}
\ee
In order to find the set of two- and zero-quark operators in the
lattice theory one has to classify the appropriate tensors according
to their transformation properties under the spin covering of the
hypercubic group. Following similar arguments as in the case of
four-quark operators, one is left with only one operator, namely
\be
   {\cal Q}_2^{\pm,\rm bare} = (m_u^2-m_c^2)\Big\{
      m_d({\sbar}P_{+}\tilde{d})
     +m_s({\sbar}P_{-}\tilde{d}) \Big\},
\ee
where $m_u,\ldots,m_c$ are the bare masses that appear in the lattice
action. 

It has been pointed out that the infinitesimal chiral transformations
in eq. (\ref{eq_InfChirTrans}) do not commute with
CP\,\cite{PHas_lat01}. This would imply that the operators in the
basis have simple transformation properties either under the chiral
symmetry or under CP, but not under both symmetries simultaneously.
The discussion in \cite{notes:4quark}, which is summarised in
Appendix~\ref{app_CP}, shows, however, that simple CP transformation
properties are recovered if one considers insertions of
${\cal{Q}}_1^{\pm,\rm bare}$ and ${\cal{Q}}_2^{\pm,\rm bare}$ in
correlation functions of local operators at non-zero distances. In
this situation, the operators transform under CP like in the continuum
theory, up to an overall factor that depends on the bare quark
masses. This is perfectly adequate for the study of issues like
operator mixing.

The upshot of this discussion is that the use of Ginsparg-Wilson
fermions yields an operator basis in the lattice-regularised theory,
whose mixing patterns are exactly like those found in the continuum.
In particular, the GIM cancellation of contributions proportional to
two-quark operators is quadratic in the masses, and thus the
coefficients that quantify the mixing of ${\cal{Q}}_1^{\pm,\rm{bare}}$
with $({\sbar}P_{\pm}\tilde{d})$ cannot develop any power divergences,
which are so hard to control for ordinary Wilson fermions.

The derivation of the effective weak Hamiltonian and the discussion of
operator mixing in the continuum (see Sect. \ref{sec_Hw_cont}) assumes
that a regularisation which preserves chiral symmetry exists. Lattice
QCD with Ginsparg-Wilson fermions is, in fact, the only known such
regularisation, so that the above discussion establishes the findings
in the continuum theory in a rigorous manner.

%% file: sect4.tex
\section{Renormalisation group invariant formulation \label{sec_RGI}}

We are now in a position to specify the effective weak Hamiltonian,
which is given as
\be
  {\cal{H}}_{\rm w}=\frac{g^2_{\rm{w}}}{4M_W^2}(V_{us})^*V_{ud}
  \sum_{\sigma=\pm}\left\{ 
       k_1^{\sigma}{\cal{Q}}_1^{\sigma}
      +k_2^{\sigma}{\cal{Q}}_2^{\sigma} \right\}.
\label{eq_Hw_final}
\ee
Here it is assumed that the operators ${\cal{Q}}_1^\pm$,
${\cal{Q}}_2^\pm$ are renormalised in a particular scheme and at a
given value of the renormalisation scale. The dependence on the scheme
and scale can be eliminated by passing to the so-called
renormalisation group independent (RGI) normalisation, for which
correlation functions of the operators stay unchanged along the
renormalised trajectory. This requires knowledge of the anomalous
dimensions, which, in the case of ${\cal{Q}}_1^\pm$, have been
determined to two loops in perturbation theory for schemes like
dimensional reduction (DRED) \cite{Alta_etal}, 't\,Hooft-Veltman (HV)
\cite{BurasWeisz90}, naive dimensional regularisation (NDR)
\cite{BurasWeisz90}, as well as the regularisation-independent (RI)
scheme \cite{Ciuch_etal,BuMisUr}. To be more explicit, consider
${\cal{Q}}_1^\pm$ in some renormalisation scheme (e.g. RI for
definiteness) at scale $\mu$. The RGI counterpart of ${\cal{Q}}_1^\pm$
is obtained via
\be
  ({\cal{Q}}_1^\pm)_{\rm RGI} = c^\pm(\mu/\Lambda)\,{\cal{Q}}_1^\pm(\mu),
\label{eq_Q1RGI}
\ee
where 
\be
   c^\pm(\mu/\Lambda)= (2b_0\gbar^2(\mu))^{\gamma_0^\pm/(2b_0)}\,
   \exp\left\{ -\int_0^{\gbar(\mu)}dg
                \left[\frac{\gamma^\pm(g)}{\beta(g)}
                      +\frac{\gamma_0^\pm}{{b_0}g} \right]\right\},
\ee
and $\gbar(\mu)$ denotes the QCD running coupling constant at scale
$\mu$. Here $\gamma^\pm(g)$ and $\beta(g)$ are the anomalous dimension
of ${\cal{Q}}_1^\pm$ and the RG $\beta$-function in the chosen scheme,
with the respective one-loop coefficients $\gamma_0^\pm$ and
$b_0$.\footnote{There is no standard convention for the overall
normalisation of RGI operators in the literature. The one adopted here
is similar to that used in \cite{alpha:mbar1} but differs, for
instance, from the commonly used normalisation for the RGI kaon
$B$-parameter.} Some perturbative results for the one-- and two--loop
coefficients are summarised in appendix~\ref{app3}.

The renormalisation of four-quark operators for lattice actions which
preserve chiral symmetry has been studied by a number of authors, both
in perturbation theory~\cite{AokiKura,SteLeo00,SteLeo01,DeGrand02},
and also non-perturbatively \cite{RBC_kpipi,BosMarBK}. For example, in
ref.~\cite{SteLeo01} the renormalisation factors ${\cal{Z}}_1^\pm$,
which relate matrix elements of ${\cal{Q}}_1^{\pm,\rm{bare}}$ in the
RI scheme to those computed using overlap fermions, were determined in
one-loop lattice perturbation theory. The RGI matrix elements can then
be obtained via the anomalous dimensions, which are known to two
loops~\cite{Ciuch_etal,BuMisUr}.

The coefficients $k_1^\pm$, $k_2^\pm$ can be considered for the RGI
operators or for those in a given continuum scheme. In either case the
coefficients $k_1^\pm$ have a well-defined perturbative expansion. The
coefficients $k_2^\pm$, though, remain undetermined, because the
effective Hamiltonian is derived from the fundamental theory by
matching a set of on-shell amplitudes, in which the total momentum
that flows into the interaction vertex vanishes, while $k_2^\pm$
multiplies a term which can be written as a total four-divergence. It
should be noted, however, that ${\cal{Q}}_2^{\pm}$ does not contribute
to the physical $K\to\pi\pi$ amplitude.

If one adopts the RGI scheme for ${\cal{Q}}_1^\pm$ according to
eq.~(\ref{eq_Q1RGI}) then the perturbative expression for $k_1^\pm$ is
obtained as
\be
   k_1^\sigma(M_W)=
   (2b_0\gbar^2(M_W))^{-\gamma_0^\sigma/(2b_0)}
   \left[1+k^{\sigma;(1)}\gbar^2(M_W)+\rmO(\gbar^4)\right],\quad\sigma=\pm.
\ee
If one follows the conventions of the $\MSbar$-scheme to define the
coupling $\gbar$, then the expression for the coefficient
$k^{\sigma;(1)}$ reads
\bea
  k^{\sigma;(1)}&=&\frac{1}{(4\pi)^2}
  \left.\frac{(\Nc-\sigma)}{4\Nc(11\Nc-2\Nf)^2} \right\{
  -693\Nc+126\Nf \nonumber\\
& & +\sigma\left.\left[
  1229\Nc^2+1881-\Nf\left(418\Nc+\frac{450}{\Nc}\right)
  +80\Nf^2\right] \right\},
\eea
where we have made explicit the dependence on the number of colours,
$\Nc$, and the number of active quark flavours, $\Nf$.

In the literature one often finds discussions of the short-distance
contribution to the $\Delta{I}=1/2$ enhancement. Since
${\cal{Q}}_1^{-}$ mediates $\Delta{I}=1/2$ transitions exclusively,
one considers the ratio
\be
   R(\mu) =
   \frac{k_1^{-}(M_W)\,k_1^{+}(\mu)}{k_1^{+}(M_W)\,k_1^{-}(\mu)},
   \qquad m_c\leq\mu\leq M_W.
\ee
The effects of physics above the charm quark mass can be estimated by
computing $R(m_c)$ via an integration of the perturbative
$\beta$-function for the coupling. One then obtains $R(m_c)\simeq2$,
which is often regarded as a ``first step'' in the explanation of the
$\Delta{I}=1/2$ rule. It must be kept in mind, however, that this
analysis is not on a solid footing, since it relies on the
applicability of two-loop perturbation theory down to scales where
corrections are of order 100\%.

%% file: sect5.tex
\section{Weak interactions in finite-volume Chiral Perturbation Theory
\label{sec_ChPT}}

We now turn to the discussion of the $\Delta{S}=1$ weak interactions
in the framework of Chiral Perturbation Theory (ChPT). In particular,
we will list expressions of correlation functions involving the
counterparts of the operators ${\cal Q}_1^\pm$, ${\cal Q}_2^\pm$ in
the effective low-energy description.

\subsection{Weak Hamiltonian in the SU(4) chiral effective theory
\label{sec_eff_weak}} 

If one considers the unphysical situation of a light charm quark, QCD
can be described at low energies by an effective Lagrangian which
possesses an $\rm SU(4)_L\times SU(4)_R$ symmetry. At leading order it
is given by
\be
  {\cal L}_{\rm E} = \quarter F^2\,\Tr\left[
  (\partial_\mu U)\partial_\mu U^\dagger \right]
  -\half\Sigma\,\Tr\left[ UM^\dagger\rme^{i\theta/\Nf}
  +MU^\dagger\rme^{-i\theta/\Nf} \right],
\label{eq_Leff_LO}
\ee
where $U\in\rm SU(4)$ denotes the field of Goldstone bosons, $\theta$
is the vacuum angle, and $M$ is the quark mass matrix. Although we are
dealing with the SU(4)-symmetric case, we have explicitly indicated
the $\Nf$-dependence in the phase factor. At this order two effective
coupling constants (``low-energy constants'', LECs), $F$ and $\Sigma$,
appear, which denote the pion decay constant and the chiral condensate
in the chiral limit, respectively. In order to incorporate the weak
interactions, we need to find low-energy transcriptions of the
operators which appear in the weak Hamiltonian,
eq.~(\ref{eq_Hw_final}), in terms of the field~$U$.

To this end we introduce the left-handed current in the low-energy
theory as
\be
  {\cal{J}}_\mu^a \equiv \half
  F^2(T^a)_{\alpha\beta}\left(
  U\partial_\mu U^\dagger\right)_{\beta\alpha},
\label{eq_Jmu_exp}
\ee
where the matrices $T^a$ denote the (hermitian) generators of
SU(4)-flavour. The current ${\cal{J}}_\mu^a$ is formally obtained by
promoting the partial derivative $\partial_\mu$ in
eq.~(\ref{eq_Leff_LO}) to a covariant one via
\be
   \partial_\mu U\longrightarrow D_\mu U=
   (\partial_\mu +iA_\mu^a T^a)U,
\ee
where $A_\mu^a$ is an external, left-handed flavour gauge field, such
that\footnote{Note that with this definition the current is formally
imaginary.}
\be
   {\cal{J}}_\mu^a = -i\left(
   \frac{\delta{\cal L}_{\rm E}}{\delta A_\mu^a}
   \right)_{A_\mu^a=0}.
\label{eq_Jmu_def}
\ee
By following a similar procedure in the fundamental theory it is easy
to see that ${\cal{J}}_\mu^a$ is the low-energy counterpart of
\be
   J_\mu^a = \psibar_\alpha\gamma_\mu
   P_{-}(T^a)_{\alpha\beta}\psi_\beta.
\ee
Furthermore, by writing the expression for the four-quark operator
$[{O}_1]_{\alpha\beta\gamma\delta}$ in \eq{eq_O1} in terms of products
of currents $J_\mu^a$, we can determine its representation in the
low-energy theory (as far as symmetries are concerned) as
\be
  [{\cal O}_1]_{\alpha\beta\gamma\delta} = 
  \quarter F^4
  \left(U\partial_\mu U^\dagger\right)_{\gamma\alpha}
  \left(U\partial_\mu U^\dagger\right)_{\delta\beta}.
\label{eq_O1_ChPT}
\ee
At leading order in the chiral expansion one can show that this is the
only operator with the same symmetry properties as its counterpart in
the full theory. In a similar manner one obtains a representation of
the two-quark operator $[{O}_2]$ as
\bea
  [{\cal O}_2]_{\alpha\beta\gamma\delta} &=& -\half\Sigma 
  \big(MM^\dagger\big)_{\gamma\beta} \Big(
   UM^\dagger \rme^{i\theta/\Nf}
  +MU^\dagger\rme^{-i\theta/\Nf} \Big)_{\delta\alpha} 
  \nonumber\\
  & & \phantom{-\Sigma}
  +(\alpha\leftrightarrow\beta, \gamma\leftrightarrow\delta).
\label{eq_O2_ChPT}
\eea
The projection of ${\cal{O}}_1$ and ${\cal{O}}_2$ onto operators
$\widehat{\cal{O}}^\pm_1$ and $\widehat{\cal{O}}^\pm_2$, which
transform under irreducible representations of dimensions 84 and 20,
is performed according to the procedure outlined in
Appendix~\ref{app1}. The effective weak Hamiltonian at leading order
in ChPT is then given by
\be
   {\cal H}_{\rm w}^{\rm ChPT} =
   \frac{g^2_{\rm{w}}}{2M_W^2}(V_{us})^*V_{ud}
   \sum_{\sigma=\pm}\left\{g_1^\sigma c_{\alpha\beta\gamma\delta}
   [\widehat{\cal{O}}_1^\sigma]_{\alpha\beta\gamma\delta}
                          +g_2^\sigma c_{\alpha\beta\gamma\delta}
   [\widehat{\cal{O}}_2^\sigma]_{\alpha\beta\gamma\delta}\right\},
\label{eq_HwChPT}
\ee
where $g_1^\pm,\,g_2^\pm$ are low-energy constants and the
coefficients $c_{\alpha\beta\gamma\delta}$ are Clebsch-Gordan-type
numbers. For the physical flavour assignments chosen in
eqs.~(\ref{eq_Q1pm}) and~(\ref{eq_Q2pm}), their values are given by
\be
   c_{suud}=1,\qquad c_{sccd}=-1,
\ee
with all other coefficients set to zero.

At leading order in ChPT the ratio of amplitudes corresponding to
$\Delta{I}=1/2$ and 3/2-transitions is given by
\be
   \frac{A_0}{A_2} = \frac{1}{\sqrt{2}}\left(
   \frac{1}{2}+\frac{3}{2}\frac{g_1^{-}}{g_1^{+}} \right).
\label{eq_A0A2_ChPT}
\ee
A non-perturbative determination of the ratio $g_1^{-}/g_1^{+}$ thus
yields direct information on the $\Delta{I}=1/2$ enhancement. In the
following we describe how the low-energy constants $g_1^\pm$ and
$g_2^\pm$ can be computed by matching suitable correlation functions
evaluated in QCD to their analytic expressions obtained in ChPT.

\subsection{Chiral Perturbation Theory in finite volume}

Our task of determining the low-energy constants in
eq.~(\ref{eq_HwChPT}) through a comparison of correlators evaluated in
lattice QCD with the expressions of ChPT can only succeed if the
latter are valid in the parameter range accessible by lattice
simulations. Numerical simulations of lattice QCD are necessarily
performed in a finite volume, and therefore the approach to the chiral
limit, where the predictions of ChPT are most accurate, will
inevitably lead to strong finite-volume effects. However, within the
framework of ChPT it is possible to account for finite-size effects in
a systematic manner. Although our actual simulations, which we
describe in Sect.~\ref{sec_num}, do not yet reach the limit of very
small quark masses, we nevertheless review here the theoretical
considerations relevant for that situation, given that reaching this
regime plays a central r\^ole for the general strategy presented in
this paper.

It was realised by Gasser and Leutwyler \cite{GasLeut_eps} (see also
\cite{Neuberger_eps}) that, in a finite volume, low-momentum modes
become increasingly important as the quark masses approach the chiral
limit. Through a re-definition of the chiral counting rules one
defines the so-called $\epsilon$-expansion, which represents a
systematic low-energy description of QCD in a finite volume for
arbitrarily small quark masses. Accordingly, the field $U\in\rm SU(4)$
is written as
\be
   U(x)=\rme^{i2\xi(x)/F}\,U_0,
\ee
where $\xi$ describes non-zero momentum modes only, while $U_0$ is a
constant SU(4) matrix collecting the zero modes. The integration over
$U_0$ must be carried out exactly when
$m\Sigma{V}\;\lesssim\;\rmO(1)$, where $m$ is the quark mass, while
the integration over the non-zero modes may be carried out
perturbatively, provided that $F$ in units of the box size is large:
$F L\gg1$. The power counting rules for the $\epsilon$-expansion are
then
\be
   F\sim\rmO(1),\quad \partial_\mu\sim\rmO(\epsilon),\quad
   L_\mu\sim\rmO(1/\epsilon),\quad \xi\sim\rmO(\epsilon),\quad
   m\sim\rmO(\epsilon^4),
\ee
where $L_\mu$ is the box length in direction $\mu$. Note that the
quark mass counts as four powers of momenta, rather than two, as in
standard ChPT in infinite volume. This has the important consequence
that additional interaction terms, which appear beyond leading order,
are suppressed if they contain one or more powers of the quark
mass. For instance, if one works at next-to-leading order, including
corrections of $\rmO(\epsilon^2)$, the physical pion mass and decay
constant, $M_\pi$ and $F_\pi$, differ from their leading order values
only by terms of relative order $\epsilon^4$ \cite{HanLeut91}, owing
to the fact that no higher-order terms contribute to the action at
order $\epsilon^2$. Furthermore, no additional interaction terms are
generated in the effective weak Hamiltonian at order $\epsilon^2$, and
hence the knowledge of the associated LECs is not required in the
analysis of the $\Delta{I}=1/2$-rule \cite{HerLai02}. The fact that
higher-order terms do not contribute at next-to-leading order in the
$\epsilon$-regime represents an enormous simplification over the
standard chiral expansion, where a large number of additional
interaction terms arises at next-to-leading order in the effective
weak Hamiltonian\,\cite{KamMissWyl90}.

The LECs $g_1^\pm,\,g_2^\pm$ in the effective weak Hamiltonian can
then be determined via the matching of correlation functions in the
$\epsilon$-regime, i.e. in a finite volume, close to the chiral
limit. Such a procedure has already been applied with some success for
quantities such as the quark condensate
$\Sigma$\,\cite{HJL99,DeGrand_cond,Bern02} and the pion decay constant
\cite{PP,lma}.

\subsection{Correlators in the $\epsilon$-regime \label{sec_corrs_eps}}

We now proceed to define the relevant correlation functions. Details
of the calculation are presented in Appendix~\ref{app4}. We also refer
the reader to ref.~\cite{HerLai02}, which gives a detailed account of
the same calculation in the more conventional $\rm SU(3)_L\times
SU(3)_R$ case.

We focus on correlation functions of two left-handed currents at
Euclidean times $x_0$ and $y_0$, respectively, and operators
$\widehat{\cal O}_1^\pm$, $\widehat{\cal O}_2^\pm$ at $z=0$. Their
definitions are
\bea
   {\cal C}^{ab}(x_0) &=& \int\rmd^3x \left\langle{\cal J}^a_0(x)
   {\cal J}^b_0(0)\right\rangle, \label{eq_Cab_def}  \\
   {[\widehat{\cal C}_1^\pm(x_0,y_0)]}^{ab}_{\alpha\beta\gamma\delta}
   &=& \int\rmd^3x\int\rmd^3y
   \left\langle
   {\cal J}^a_0(x)[\widehat{\cal O}_1^\pm(0)]_{\alpha\beta\gamma\delta}
   {\cal J}^b_0(y)
   \right\rangle, \label{eq_C1pm_ChPT}  \\
   {[\widehat{\cal C}_2^\pm(x_0,y_0)]}^{ab}_{\alpha\beta\gamma\delta}
   &=& \int\rmd^3x\int\rmd^3y
   \left\langle
   {\cal J}^a_0(x)[\widehat{\cal O}_2^\pm(0)]_{\alpha\beta\gamma\delta}
   {\cal J}^b_0(y)
   \right\rangle,
\eea
where the integrations are performed over the spatial volume
$L_1L_2L_3$. Since the projected operators
$[\widehat{\cal{O}}_1^\pm]_{\alpha\beta\gamma\delta}$ are linear
combinations of $[{\cal{O}}_1]_{\alpha\beta\gamma\delta}$, the
correlation function
$[\widehat{\cal{C}}_1^\pm]^{ab}_{\alpha\beta\gamma\delta}$ is obtained
as a sum of correlators of the individual operators ${\cal{O}}_1$. The
same holds for
$[\widehat{\cal{C}}_2^\pm]^{ab}_{\alpha\beta\gamma\delta}$.

\begin{figure}
\begin{center}
\psfig{file=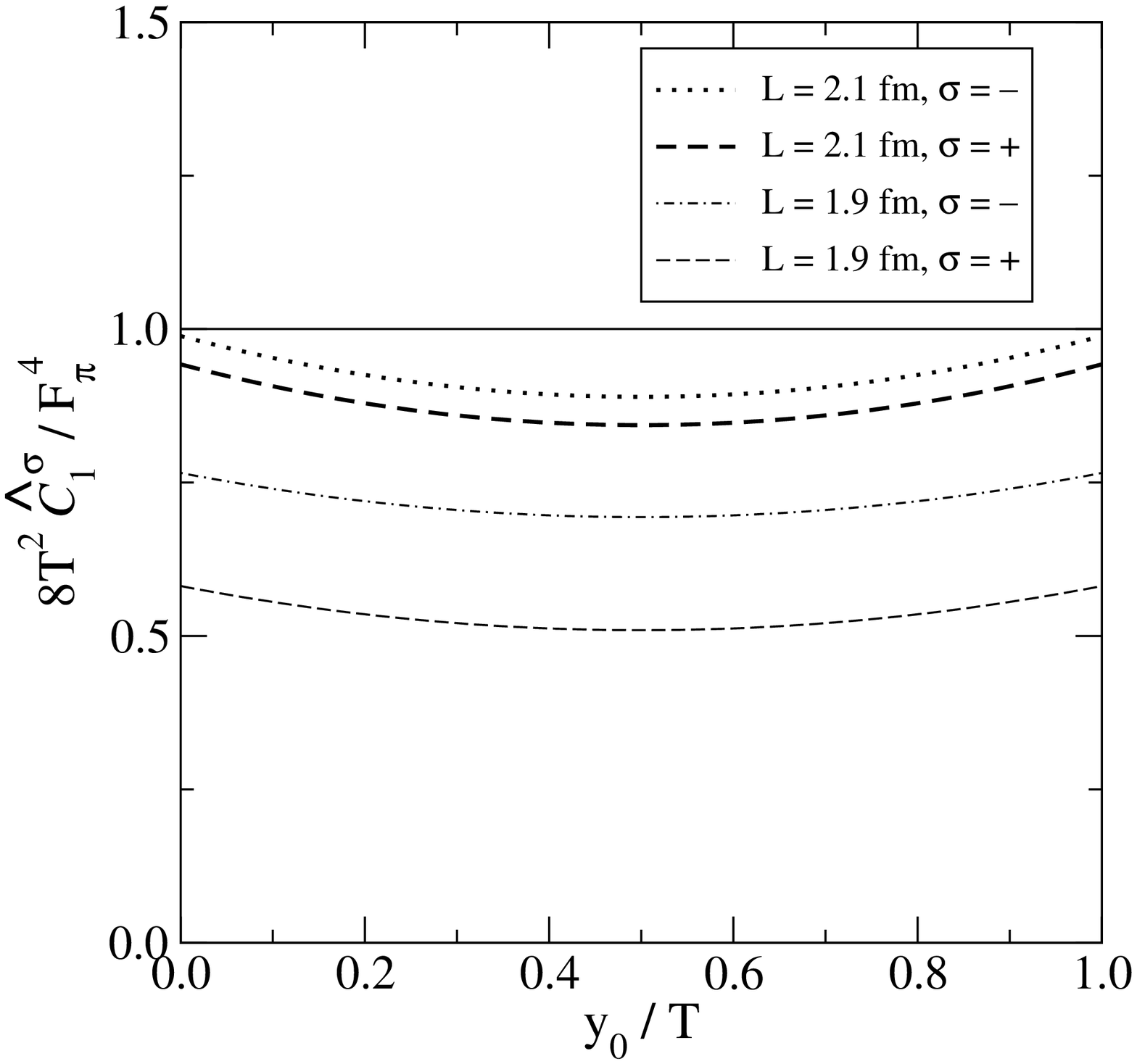,width=7.cm}
\psfig{file=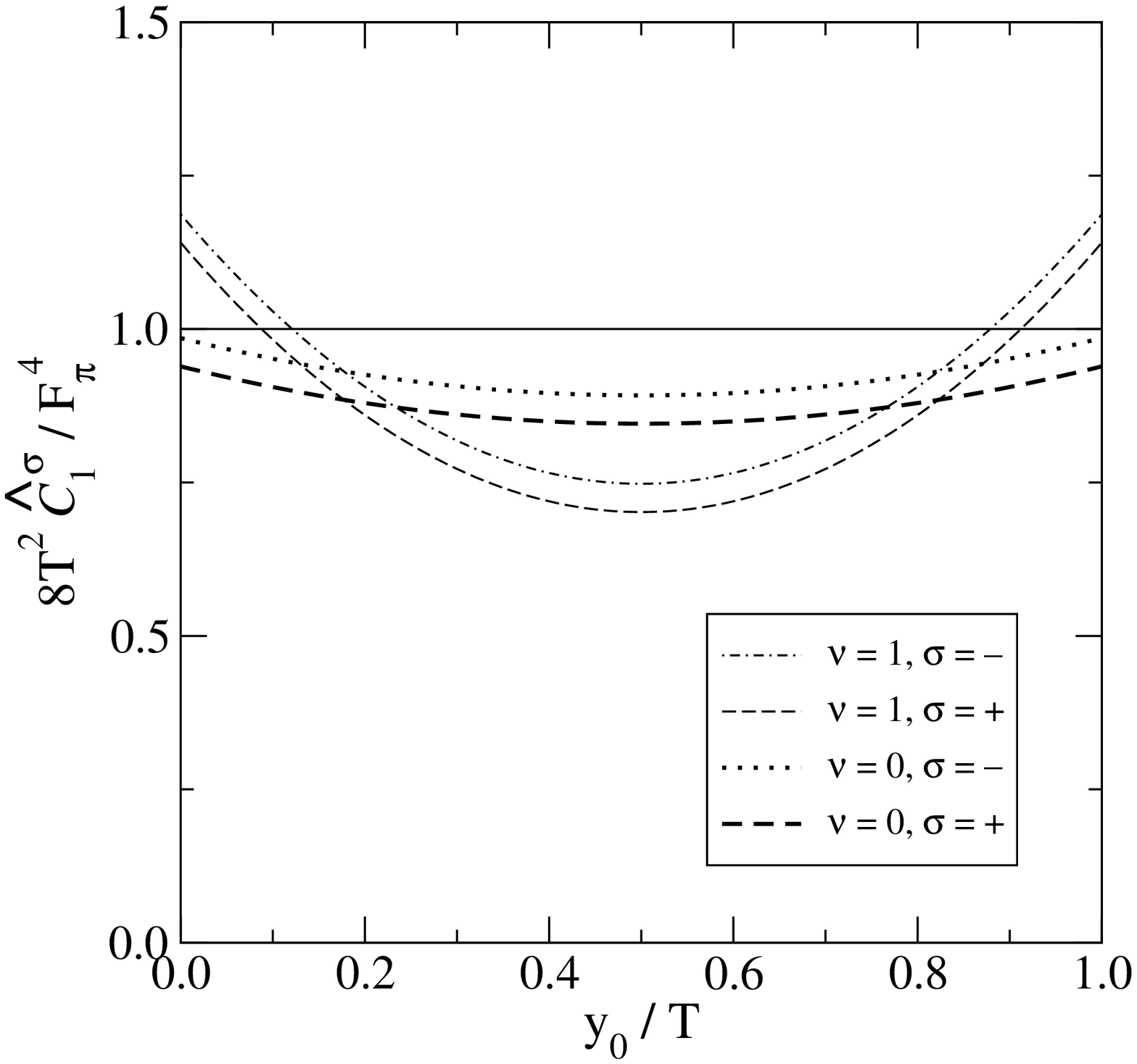,width=7.cm}
\caption{\footnotesize Left: the correlator
$\widehat{\cal{C}}_1^\sigma(x_0,y_0)$ of \protect\eq{eq_C1_res},
normalised to the tree-level value (solid line). The time extent is
$T=3.0\,\fm$, the quark mass $m = 5\,\MeV$, the $x_0$-coordinate
$x_0/T = - 1/4$, the pion decay constant $F_\pi = 93\,\MeV$, and the
chiral condensate $\Sigma=(250\,\MeV)^3$. Right: the same observable
with the spatial extent $L = 2.1\,\fm$, but in an ensemble with a
fixed topological charge $\nu$. \label{fig_ChPT_3pt}}
\end{center}
\end{figure}

The general expressions for the correlation functions are presented in
Appendix~\ref{app4} and\,\cite{HerLai02}. Here we only quote the
results for a specific set of parameters, which corresponds to the
unphysical situation of a light charm quark. We choose a diagonal mass
matrix, $M={\rm{diag}}(m,m,m,m)$, and work in a
four-volume $V=T\times L^3$. Keeping in mind that $F_\pi=F$ and
$M_\pi=(2m\Sigma/F^2)^{1/2}$ at order $\epsilon^2$, we find
\be
  {\cal C}^{ab}(x_0)=\Tr(T^a T^b)\frac{F_\pi^2}{2T}\left\{
  1+\frac{1}{F_\pi^2T^2}\rho^3\Big[\Nf(\beta_1\rho^{-3/2}-k_{00})
  +uC_\theta(u/2)\,h_1(x_0/T)\Big]
  \right\},
\label{eq_Cab_res}
\ee
where $\rho=T/L$, and $C_\theta(u/2)$ is defined in
\eq{eq_Ctheta_Cnu}. Here we keep explicit factors of $\Nf$, as this
will turn out to be useful when the quenched theory is considered
later. The function $h_1$ and the ``shape coefficients'' $\beta_1$ and
$k_{00}$ are specified in Appendix~\ref{app4}. The variable $u$ is
given by $u=M_\pi^2 F_\pi^2
V=2m{\Sigma}V$~\cite{GasLeut_eps,Hansen90,HanLeut91}.
For the same parameter choice, and assuming that the generators $T^a$
and $T^b$ are chosen according to \eq{eq_TaTb}, we find for
$\widehat{\cal{C}}_1^{\pm}(x_0,y_0)$:
\bea
   \widehat{\cal C}_1^\sigma(x_0,y_0) &=& \frac{F_\pi^4}{8T^2}
   \bigg\{1 +\frac{1}{F_\pi^2T^2}\rho^3
     \Big[ 2(\Nf+\sigma)(\beta_1\rho^{-3/2}-k_{00})
   \label{eq_C1_res}  \\ 
       & &
\phantom{-\frac{F_\pi^4}{8T^2}\bigg\{1+\frac{1}{F_\pi^2T^2}}
          +uC_\theta(u/2)\Big(h_1(x_0/T)+h_1(y_0/T)\Big) \Big]\bigg\},
   \quad\sigma=\pm  \nonumber
\phantom{FF}
\eea
where we have suppressed flavour indices for brevity. In order to
study the convergence properties at this order of the expansion, we
plot $\widehat{\cal C}_1^\sigma(x_0,y_0)$ in
Fig.~\ref{fig_ChPT_3pt}\,(left panel) for a volume with time extent
$T=3.0\,\fm$ and two different spatial box sizes $L$, corresponding to
1.9 and 2.1\,\fm, respectively. We have also set $\theta=0$ here. One
clearly sees that the corrections at order $\epsilon^2$ are quite
large for $L\;\lesssim\;2.0\,\fm$.

The above results are obtained in a fixed $\theta$-vacuum. By
performing a Fourier transform in $\theta$ one can obtain averages in
sectors of ``fixed topology'', characterised by an index $\nu$. In
ref.\,\cite{LeutSmil92} the special r\^ole of topology in the
$\epsilon$-regime was emphasised, noting that certain quantities
depend quite strongly on the index near the chiral limit. Recent
developments in the understanding of the r\^ole of topology show that
correlation functions in fixed topological sectors can be defined also
in QCD \cite{topol1,topol2}. Although this result is universal
\cite{topol2}, different topological sectors can be identified through
the index of the fermion operator only if fermionic discretisations
with an exact chiral symmetry are used.

When considering averages in sectors of fixed $\nu$, poles in the
quark mass $m$ are expected to occur in certain correlation
functions. Whether or not such poles also appear in quantities
considered here is what we discuss in the following. Note that in the
regime of larger quark masses (conventionally called $p$-regime),
topology does not play a major r\^ole, and it is commonplace to
present predictions of ChPT for fixed $\theta$ rather than $\nu$.

The transition to sectors of fixed topology simply amounts to
substituting $C_\theta(u/2)$ by $C_\nu(u/2)$, whose definition is
specified in Appendix~\ref{app4}. As can be seen from
Fig.\,\ref{fig_ChPT_3pt}\,(right panel) the time behaviour of the
correlator is indeed strongly modified in the small-mass
region. Although the correlators do not develop any poles for $m\to0$
(because $C_\nu(u/2)$ is multiplied by $u$), their time dependence
does not vanish in this limit, for $\nu \neq 0$.

So far we have ignored the correlation function
$[\widehat{\cal{C}}_2^\pm]$. The reason is that it arises only at
order $\epsilon^8$, owing to the fact that it contains two powers of
the mass matrix each contributing a factor of order
$\epsilon^4$. Actually, if a diagonal mass matrix and flavour
assignments as in the derivation of \eq{eq_C1_res} are chosen,
$[\widehat{\cal C}_2^\pm]$ vanishes identically. In order to isolate
the ``pure QCD'' contribution to the $\Delta{I}=1/2$ enhancement at
next-to-leading order, it is thus sufficient to concentrate on the
LECs $g_1^\pm$, dropping the operator
$[\widehat{\cal{O}}_2^\pm]_{\alpha\beta\gamma\delta}$ altogether.

\begin{figure}
\begin{center}
\psfig{file=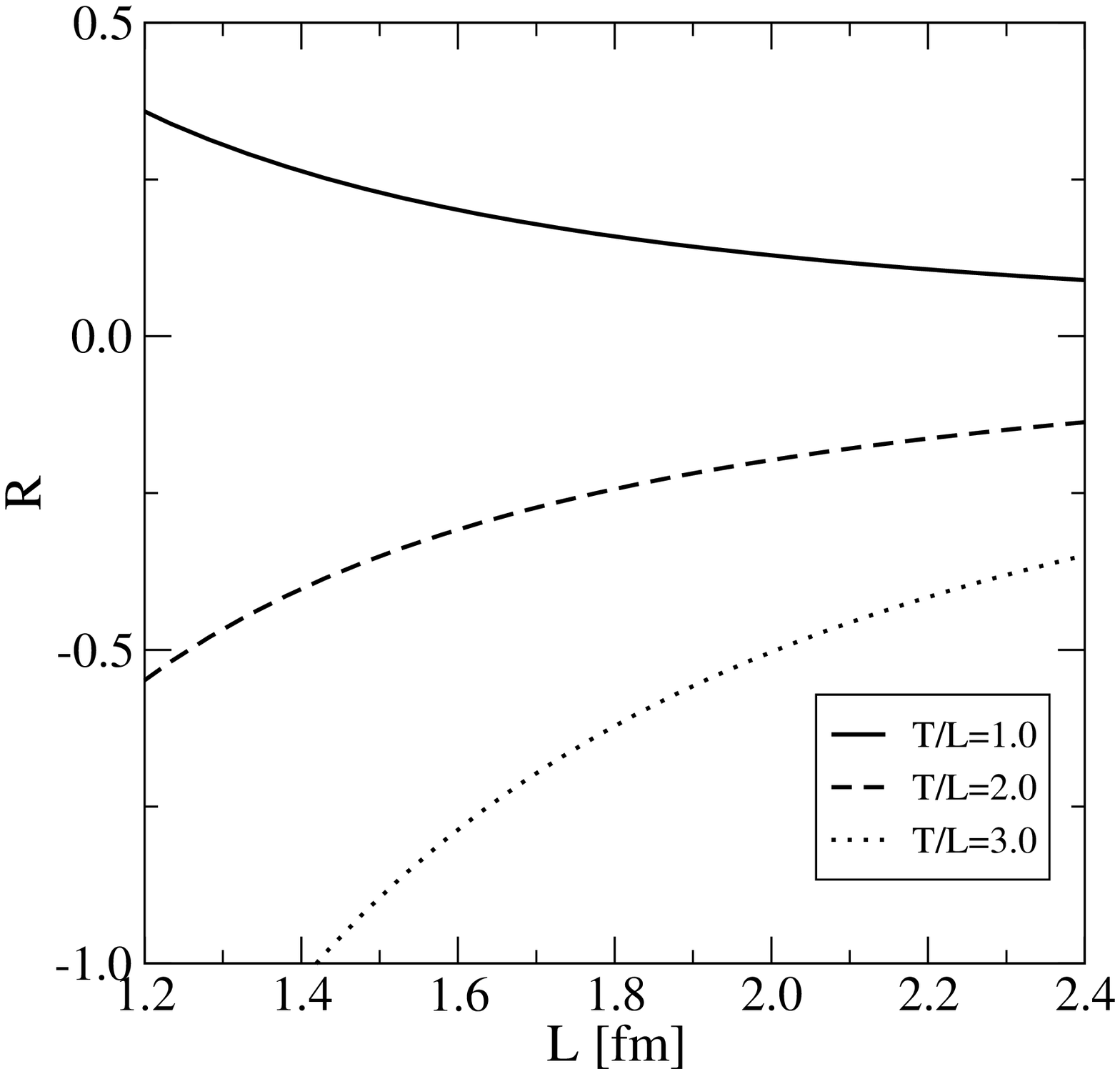,width=7.cm}
\psfig{file=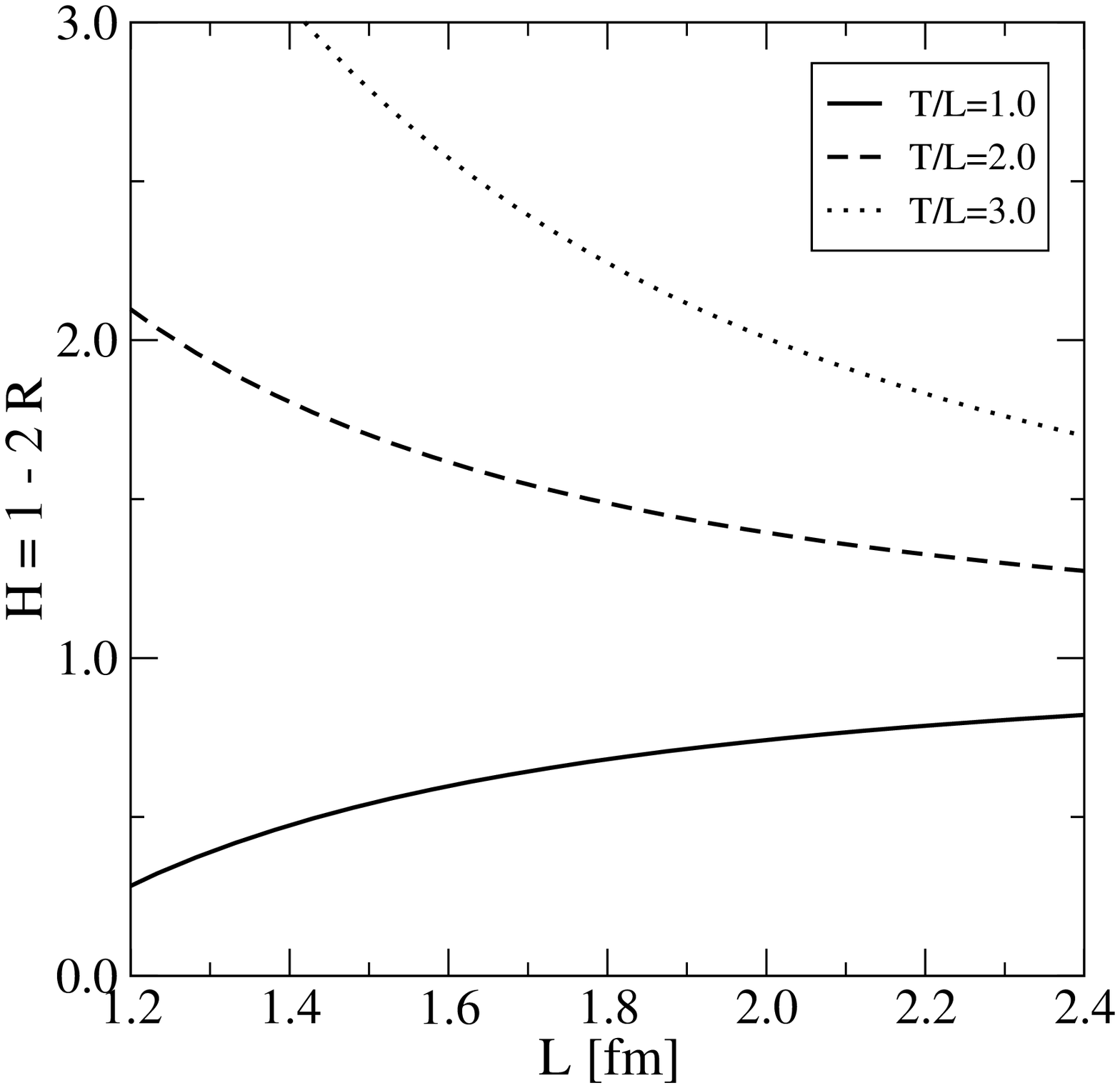,width=7.cm}
\caption{\footnotesize The quantities $R(x_0,y_0)$ (left panel) and
$H(x_0,y_0)$ (right panel) plotted as functions of the box size and
geometry. For $F$ we have here chosen a value indicated by quenched
simulations at modest volumes, $F\approx103\,\MeV$ \cite{lma}, but the
dependence of $R$ on $F$ is only through an overall factor $1/(FT)^2$
and the curves have thus the same form for any other value as well.
\label{fig_3pt_rat}}
\end{center}
\end{figure}

The determination of $g_1^{+}$ and $g_1^{-}$ is greatly facilitated by
considering suitable ratios of correlation functions. If we define
\be
   K^{\pm}(x_0,y_0) \equiv
   2\frac{\widehat{\cal C}_1^{\pm}(x_0,y_0)}
          {{\cal C}^{a a^\dagger}(x_0){\cal C}^{b b^\dagger}(y_0)},
\label{eq_Kpm_def}
\ee
choose
$T^{a^\dagger}\equiv(T^a)^\dagger$, $T^{b^\dagger}\equiv(T^b)^\dagger$,
with $T^a,\,T^b$ as in \eq{eq_TaTb}, and insert the expressions
derived above we obtain
\be
   K^{\sigma}(x_0,y_0) =
   1+\frac{2\sigma}{F_\pi^2 T^2}\rho^3\left\{
   \beta_1\rho^{-3/2}-k_{00} \right\},\quad\sigma=\pm.
\label{eq_Ksigma}
\ee
Thus, at order $\epsilon^2$ the dependence on $m$, $x_0$, $y_0$ and
--- if the correlations are computed for fixed topology --- also on
$\nu$ drops out in the ratios $K^{\pm}$. If we write
\be
   K^\sigma(x_0,y_0)=1+\sigma R(x_0,y_0),\quad\sigma=\pm,
\ee
we can plot the expression for
$R(x_0,y_0)$ (which is independent of $x_0,y_0$ at order $\epsilon^2$)
as a function of the volume, see Fig.~\ref{fig_3pt_rat}.
Another important ratio is
\be
   H(x_0,y_0) \equiv
  \frac{\widehat{\cal C}_1^{-}(x_0,y_0)}
       {\widehat{\cal C}_1^{+}(x_0,y_0)}
  = \frac{K^{-}(x_0,y_0)}{K^{+}(x_0,y_0)}.
\ee
In terms of the one-loop correction $R(x_0,y_0)$ it is given at
$\rmO(\epsilon^2)$ by
\be
   H(x_0,y_0) = 1-2R(x_0,y_0)
  = 1-\frac{4}{F_\pi^2 T^2}\rho^3\left\{
   \beta_1\rho^{-3/2}-k_{00} \right\},
\label{eq_Hexp}
\ee
and the right panel in Fig.~\ref{fig_3pt_rat} shows a plot of $H$ as a
function of the volume for different geometries. The computation of
the corresponding ratio of three-point functions in lattice QCD in the
$\epsilon$-regime is then directly proportional to $g_1^{-}/g_1^{+}$:
\be
   \frac{g_1^{-}}{g_1^{+}} = \frac{k_1^{-}}{k_1^{+}}
   \frac{C_1^{-}(x_0,y_0)}{C_1^{+}(x_0,y_0)}
   \frac{1}{H(x_0,y_0)}, 
\label{eq_g1pm_R2}
\ee
where
\be
   C_1^{\pm}(x_0,y_0) = \sum_{\xvec,\yvec}\left\langle
   J_0(x)\,{\cal Q}_1^{\pm}(0)\,J_0(y) \right\rangle
\ee
is the three-point correlation function in QCD. Here it is assumed
that the Wilson coefficients $k_1^\pm$ are known for the particular
scheme in which the operators ${\cal Q}_1^{\pm}$ are renormalised.

Let us stress that, in the main result of this section, \eq{eq_Hexp},
the Goldstone zero-momentum mode contribution, $C_\theta(u/2)$ or
$C_\nu(u/2)$, has cancelled completely at this order. This means that
the result in the $\epsilon$-regime at NLO can be obtained from a NLO
calculation of the same quantity in the $p$-regime, by taking just the
leading term in a Taylor expansion in the pseudoscalar mass
$\mps^2\equiv 2 m \Sigma/F^2$. Note that, as long as the volume is
kept fixed, the contributions from non-zero modes are infrared-safe,
since the chiral logarithms encountered in infinite volume are
replaced by logarithms of the volume. The fact that the expression for
the ratio $\widehat\mathcal{C}_1^-/\widehat\mathcal{C}_1^+$ in the
$p$-regime can be Taylor-expanded in $\mps^2$ for fixed volume can also
be used in practice to obtain the result in the $\epsilon$-regime
indirectly from numerical simulations in the $p$-regime, by fitting
the numerically determined ratio
$\widehat\mathcal{C}_1^-/\widehat\mathcal{C}_1^+$ to a constant (which
is the result in the $\epsilon$-regime) plus ${\rmO}(\mps^2)$
corrections with some unknown coefficient.

\subsection{The quenched approximation}

Currently, numerical simulations of lattice QCD with Ginsparg-Wilson
fermions are mostly restricted to the quenched approximation, owing to
the large computational cost. Here we discuss the implications of
quenching for our strategy.

Quenched QCD can be treated within the framework of an effective
Lagrangian, with the resulting low-energy expansion referred to as
quenched Chiral Perturbation Theory (qChPT)
\cite{BandG92,Sharpe92}. The theoretical status of qChPT is, however,
questionable. This manifests itself in the occurrence of infrared
divergences in certain correlation functions. These divergences
reflect, at least partially, the sickness of quenched QCD. Here we
adopt the pragmatic assumption that --- despite the fact that it is
not an asymptotic expansion of quenched QCD (for a fixed number of
colours $\Nc$) --- quenched ChPT does describe the low-energy regime
of quenched QCD in certain ranges of kinematical scales, where
correlation functions can be parameterised in terms of effective
coupling constants, the latter being defined as the couplings which
appear in the Lagrangian of the effective theory. Under this
assumption we now discuss the determination of the counterparts of
$g_1^\pm$ in the quenched theory, via the matching of correlation
functions computed in quenched QCD and qChPT.

The main difference between ChPT considered in the quenched and
unquenched theories is that in the former there is no decoupling of
flavour singlets from the pseudo-Goldstone bosons. The dynamics of the
singlet field $\Phi_0$ must then be incorporated into the effective
chiral Lagrangian, which requires additional LECs associated with the
new interaction terms. To be more precise, we consider the quenched
effective Lagrangian in the so-called ``replica formalism''
\cite{DamSplit00,DamDiHerJa01}
\bea
  {\cal L}_{\rm E}^{\rm quen} &=& \quarter F^2\,\Tr\left[
  (\partial_\mu U)\partial_\mu U^{-1} \right]
  -\half\Sigma\,\Tr\left[ U_\theta UM^\dagger+MU^{-1}U_\theta^{-1}
   \right] \nonumber \\
  & &+\frac{m_0^2}{2\Nc}\Phi_0^2
  +\frac{\alpha}{2\Nc}(\partial_\mu\Phi_0)^2. 
\label{eq_Leff_quen}
\eea
In the replica method one introduces $\Nv$ valence quarks that are
considered independently from the number of sea quarks, $\Nf$, which
will eventually be taken to zero. In the quenched case the fields $U$
are promoted from elements of SU($\Nf$) to those of U($\Nf$). The
factor $U_\theta$ is given by $U_\theta=\exp(i\theta I_{\Nv}/\Nv)$,
where $I_{\Nv}$ is the unit matrix in the subspace corresponding to
valence quarks. In addition to the LECs $F$ and $\Sigma$ (which, of
course, may differ from those in the full theory), there are new
parameters, $\alpha$ and $m_0$, which are associated with the singlet
field $\Phi_0$.

We have noted above that poles in the quark mass may appear in fermion
propagators if the theory is considered for fixed non-trivial
topology. These poles are expected to be the same in the full and
quenched cases. For fixed topology, the counting rules of the
$\epsilon$-expansion remain unchanged when passing to the quenched
theory.

When the $\Delta{S}=1$ weak interactions are incorporated into the
quenched setting, the larger symmetry group and the presence of the
singlet field $\Phi_0$ may allow for additional interactions that need
to be taken into account at a given order of the chiral
expansion. Without going into further detail, we note that the current
${\cal J}_\mu^a$ and the operators ${\cal O}_1$ and ${\cal O}_2$ need
not be modified in the quenched case, provided that an additional
expansion in $1/\Nc$ is assumed, which is needed in order to justify
the truncation in operators involving higher powers of $\Phi_0$. This
truncation has also been invoked to arrive at
\eq{eq_Leff_quen}. Furthermore, the effective weak Hamiltonian does
not have to be supplemented by additional operators. The
classification of operators according to the valence symmetry does not
give rise to ambiguities like those discussed in \cite{GolPall} for
operators that appear in the conventional SU(3)-case. The argument
which proves the absence of these ambiguities is analogous to the one
discussed in \cite{HerLai02} for the 27-plet operator in the SU(3)
case. We refer the reader to this reference for more details.

We now give the expressions for the quenched analogues of the
correlation functions in eqs. (\ref{eq_Cab_res}) and~(\ref{eq_C1_res})
for fixed topology. They are formally obtained by taking $\Nf\to0$ in
the general expressions for the correlators obtained from the
Lagrangian in \eq{eq_Leff_quen}. For the two-point function this
yields \cite{VandA,HerLai02}
\be
  {\cal C}^{ab;\rm quen}(x_0)=\Tr(T^a T^b)\frac{F^2}{2T}\left\{
  1+\frac{1}{F^2T^2}\rho^3\Big[uC_\nu^{\rm quen}(u/2)\,h_1(x_0/T)\Big]
  \right\},
\label{eq_Cab_res_quen}
\ee
where
\be 
   C_\nu^{\rm quen}=\frac{1}{\Nv}\left\langle \Re\Tr_{\rm v} U_0
   \right\rangle_{\nu,U_0} 
\ee
and $\Tr_{\rm v}$ denotes the trace over the valence subgroup. The
precise meaning of the average $\langle\Re\Tr_{\rm v} U_0
\rangle_{\nu,U_0}$ is specified in \cite{HerLai02}. Note that there is
no dependence on the singlet couplings $m_0^2$ and $\alpha$. As a
result, it is easy to see that the quenched results can be simply
obtained from the unquenched expressions in the limit $\Nf\to0$. In a
similar manner one finds the result for the three-point function
\bea
   \widehat{\cal C}_1^{\sigma;{\rm quen}}(x_0,y_0) &=& \frac{F^4}{8T^2}
   \bigg\{1 +\frac{1}{F^2T^2}\rho^3
     \Big[ 2\sigma(\beta_1\rho^{-3/2}-k_{00}) \\
       & &
\phantom{-\frac{F_\pi^4}{8T^2}\bigg\{1+\frac{1}{F_\pi^2T^2}\rho^3\Big[}
          +uC_\nu^{\rm quen}(u/2)\Big(h_1(x_0/T)+h_1(y_0/T)\Big) \Big]
          \bigg\}.  \nonumber
\label{eq_C1_res_quen}
\phantom{FF}
\eea
One can easily see that the ratios $K^\pm(x_0,y_0)$ and~$H(x_0,y_0)$
defined above remain unchanged, and thus the relation in
\eq{eq_g1pm_R2} between the ratio $g_1^{-}/g_1^{+}$ and the
correlation functions carries over to the quenched case without
modification, at this order. The same is true for the expansion of the
ratio $\widehat{\cal{C}}_1^{-}/\widehat{\cal{C}}_1^{+}$ in $\mps^2$ at
fixed volume, which was discussed at the end of
Sect.~\ref{sec_corrs_eps}.

%% file: sect6.tex
\section{Decoupling of the charm quark in Chiral Perturbation Theory
\label{sec_charm_ChPT}}

We will now depart from the unphysical situation of a charm quark
degenerate with the other light quarks and investigate its effects in
the framework of ChPT. In particular, we will show how the
$\Delta{I}=1/2$ and 3/2 amplitudes evolve as $m_c$ is increased, but
stays below the QCD scale, in order for ChPT to remain applicable. The
aim is to find a relation between the LECs $g_1^\pm$ and the
corresponding couplings in the theory where the charm quark has been
integrated out. Here we only sketch the procedure and state the main
results: a detailed account of the calculation is given in a separate
publication~\cite{HerLai_prep}.

\subsection{Basic setup}

The effective Lagrangian ${\cal L}_{\rm E}$ of QCD with an $\rm
SU(4)_L\times SU(4)_R$ chiral symmetry was defined to leading order
in\,\eq{eq_Leff_LO}. Since the decoupling of the charm quark is
insensitive to infrared physics, we will, for simplicity, consider the
theory in infinite volume. Unlike the situation encountered in the
$\epsilon$-regime, counterterms must then be added to the effective
Lagrangian, if higher orders in the expansion are considered. For our
purposes it is sufficient to include the counterterm
\be
   \delta{\cal L}_{\rm E} = 
   L_4\,\Tr[\partial_\mu U\partial_\mu U^\dagger] 
        \Tr[\chi^\dagger U+U^\dagger\chi],
\label{eq_LE_NLO}
\ee
where the vacuum angle $\theta$ has been set equal to zero and $\chi$
is defined as
\be
   \chi=\frac{2\Sigma M}{F^2}.
\ee
Phenomenological estimates for the value of $L_4$ are available in the
physical $\rm SU(3)_L\times SU(3)_R$-symmetric
case~\cite{GasLeu85,LECs_phen}.

Weak interactions are incorporated through the effective Hamiltonian,
which, at leading order, contains the operators ${\cal O}_1$ and
${\cal O}_2$ given in eqs.~(\ref{eq_O1_ChPT})
and~(\ref{eq_O2_ChPT}). We now consider a mass matrix of the form
\be
   M={\rm diag}(m_u,m_d,m_s,m_c)
\ee
with $m_c\gg m_u=m_d=m_s$, noting that the restriction to degenerate
non-charm quarks is irrelevant for our purposes. Furthermore we define
\be
   \chi_{uu}\equiv\frac{2m_u\Sigma}{F^2},\quad
   \chi_{cc}\equiv\frac{2m_c\Sigma}{F^2}.
\ee
If momenta $p$ with $p^2\;\gtaeq\;\chi_{cc}$ are excluded, then one
expects the physics of the SU(4)-symmetric theory to be described by
another effective Lagrangian from which the heavy scale has been
integrated out, and which possesses an $\rm SU(3)_L\times SU(3)_R$
chiral symmetry. Our task is now to derive the effective weak
Hamiltonian of such a theory, assuming that the LECs $g_1^\pm$ are
already known in the SU(4)-symmetric case. In order to define our
power-counting rules we assume the following hierarchies:
\be
   \chi_{uu}\ll\chi_{cc}\ll(4\pi F)^2.
\ee
In accordance with these relations, we work at order
$\chi_{cc}/(4{\pi}F)^2$ in the chiral expansion, while dropping all
terms of order $\chi_{uu}/\chi_{cc}$ and $\chi_{uu}/(4{\pi}F)^2$.

The form of the strong interaction part of the effective $\rm
SU(3)_L\times SU(3)_R$ theory is identical to ${\cal L}_{\rm E}$, but
formulated in terms of SU(3) matrices and modified LECs. In order to
distinguish them from their counterparts in the $\rm SU(4)_L\times
SU(4)_R$-symmetric theory, we will denote them with a bar. The
effective action then reads
\be
  \bar{\cal L}_{\rm E} = \quarter \bar{F}^2\,\Tr\left[
  (\partial_\mu \bar{U})\partial_\mu \bar{U}^\dagger \right]
  -\half\bar\Sigma\,\Tr\left[ \bar{U}\bar{M}^\dagger
  +\bar{M}\bar{U}^\dagger \right],
\ee
where $\bar{U}\in\rm SU(3)$ and
\be
 \bar{M}={\rm diag}(m_u,m_d,m_s).  
\ee 
Flavour indices are denoted by $\balpha,\bbeta,\ldots$ and take their
values from the set $(u,d,s)$. The construction of non-singlet weak
operators transforming under
${\bf{3}}^*\otimes{\bf{3}}^*\otimes{\bf{3}}\otimes{\bf{3}}$ of $\rm
SU(3)_L$ proceeds along the same lines as in the SU(4)-case. By
considering the power-counting rules introduced above, one finds that
the SU(3)-counterpart of ${\cal O}_2$ is of relative order
$(\chi_{uu}/\chi_{cc})^2$ and can thus be dropped. This leaves
\footnote{In order to make contact with the conventions of
ref.~\cite{HerLai02}, we note that there the operator $\bar{\cal O}_1$
is denoted by ${\cal{O}}_w$, while our ${[\bar{\cal{O}}_3]}_{sd}$
corresponds to ${\cal{O}}_8^\prime$ of \cite{HerLai02}.}
\bea
    {[\bar{\cal O}_1]}_{\balpha\bbeta\bgamma\bdelta} &\equiv& \quarter
    \bar{F}^4
    \left(\bar{U}\partial_\mu\bar{U}^\dagger\right)_{\bgamma\balpha} 
    \left(\bar{U}\partial_\mu\bar{U}^\dagger\right)_{\bdelta\bbeta} \\
    {[\bar{\cal O}_3]}_{\balpha\bgamma} &\equiv& \half
    \bar{F}^2\bar{\Sigma}\big(\bar{U}\bar{M}^\dagger
    +\bar{M}\bar{U}^\dagger\big)_{\bgamma\balpha},
\eea
and the effective weak Hamiltonian, denoted by
$\bar{\cal{H}}_{\rm{w}}^{\rm ChPT}$, is a combination of these two
operators. The matching proceeds by enforcing equality between
correlation functions involving left-handed flavour currents. In
analogy with the procedure outlined in Sect.~\ref{sec_eff_weak}, the
latter are derived by defining covariant derivatives according to
\be
  \partial_\mu{U}\to D_\mu{U}\equiv(\partial_\mu+iA_\mu^a\bar{T}^a)U,
  \qquad
  \partial_\mu\bar{U}\to
  D_\mu\bar{U}\equiv(\partial_\mu+iA_\mu^a\bar{T}^a)\bar{U},
\ee
where $\bar{T}^a$ are the Hermitian generators of SU(3).%
\footnote{Or, more precisely, in the case of SU(4), Hermitian
generators in the sub-algebra that generates SU(3).}
The left-handed currents ${\cal{J}}_\mu^a$ and $\bar{\cal{J}}_\mu^a$
in the SU(4)- and SU(3)-symmetric theories are then obtained by taking
functional derivatives with respect to $A_\mu^a$.

In order to perform the matching to leading order in the weak
Hamiltonian we require
\bea
\left\langle{\cal J}_\mu^a(x){\cal J}_\nu^b(y)\right\rangle_{\rm SU(4)}
&=& 
\left\langle\bar{\cal J}_\mu^a(x)\bar{\cal J}_\nu^b(y)
\right\rangle_{\rm SU(3)} \label{eq_match_2pt},
\label{eq_2pt_match}\\
\left\langle{\cal J}_\mu^a(x){\cal H}_{\rm w}^{\rm ChPT}(z)
            {\cal J}_\mu^b(y)\right\rangle_{\rm SU(4)}
&=& 
\left\langle\bar{\cal J}_\mu^a(x)\bar{\cal H}_{\rm w}^{\rm ChPT}(z)
            \bar{\cal J}_\mu^b(y)\right\rangle_{\rm SU(3)},
\label{eq_match_3pt}
\eea
where the expectation values are evaluated using the strangeness
conserving Lagrangian, and it is understood that space-time
separations are large compared with the scales set by $\chi_{cc}$ and
$4{\pi}F$. The objective is then to compute the observables on the
left-hand sides of eqs.~(\ref{eq_match_2pt}) and~(\ref{eq_match_3pt})
to the order in $\chi_{cc}$ defined previously and find out how the
LECs and operators on the right-hand sides must be adjusted.

\subsection{Matching of the SU(3) and SU(4)-symmetric theories}

By enforcing the matching condition on current-current correlators one
obtains the relations between the low-energy parameters $F$ and
$\bar{F}$, as well as $\Sigma$ and $\bar\Sigma$. Here we omit all the
details of the calculation and refer to \cite{HerLai_prep}. At
one-loop level, the pion decay constant in the chiral limit in the
SU(3)-symmetric theory, $\bar{F}$, is related to $F$ via
\be
  \bar{F}^2=F^2\left\{1-\frac{1}{2}\frac{\chi_{cc}}{(4{\pi}F)^2}
  \Big[ \ln\frac{\chi_{cc}}{2\bar{\mu}^2}-1 - 16(4\pi)^2 L_4^\msbar
  \Big]\right\}.
\label{eq_F0_redef}
\ee
Here we have adopted the conventions of the $\msbar$-scheme, with
$\bar\mu$ denoting the subtraction scale. A similar relation can be
derived for the parameters $\Sigma$ and
$\bar\Sigma$~\cite{HerLai_prep}.

When matching the effective weak Hamiltonians according to
\eq{eq_match_3pt}, it is convenient to replace
${\cal{H}}_{\rm{w}}^{\rm ChPT}(z)$ by the operator ${\cal{O}}_1$ in
the $\fourby$ theory and work out the set of operators in the
$\threeby$ theory it leads to. Skipping over the details of the
calculation, we simply quote the result: when moving from
${\cal{H}}_{\rm{w}}^{\rm ChPT}$ to $\bar{\cal H}_{\rm w}^{\rm ChPT}$,
the operator $[{\cal{O}}_1]$ must be replaced by
\bea
  [{\cal{O}}_1]_{\alpha\beta\gamma\delta} &\longrightarrow& 
  c_1 [\bar{\cal{O}}_1]_{\balpha\bbeta\bgamma\bdelta}
 +c_2\Big( \delta_{{\alpha}c}\delta_{{\delta}c}
           [\bar{\cal{O}}_1]_{\bbeta\bkappa\bkappa\bgamma} 
          +\delta_{{\beta}c}\delta_{{\gamma}c}
           [\bar{\cal{O}}_1]_{\balpha\bkappa\bkappa\bdelta} \Big) \nonumber\\
  & &+d_2\Big( \delta_{{\alpha}c}\delta_{{\delta}c}
               [\bar{\cal{O}}_3]_{\bbeta\bgamma} 
              +\delta_{{\beta}c}\delta_{{\gamma}c}
               [\bar{\cal{O}}_3]_{\balpha\bdelta} \Big),
\label{eq_O1_match}
\eea
where the coefficients $c_1, c_2$ and $d_2$ are given by
\bea
& & c_1=1+2K\frac{\chi_{cc}}{F^2}, \label{eq_c1_expr}  \\
& & c_2= -\frac{3}{4}\frac{\chi_{cc}}{(4\pi F)^2}
          \ln\frac{\Lambda_\chi^2}{\chi_{cc}},\qquad
    d_2= -\frac{1}{2}\frac{\chi_{cc}}{(4\pi F)^2}
          \ln\frac{\Lambda_\chi^2}{\chi_{cc}},   \label{eq_c2d2_expr}
\eea
and SU(3) singlet structures have been omitted in \eq{eq_O1_match}.
Here $\Lambda_\chi$ is some physical scale that will in general be
different for $c_2$ and $d_2$, since it incorporates the effect of the
finite corrections (similar to the terms proportional to $L_4^\msbar$
in \eq{eq_F0_redef}, but for the weak interaction part
\cite{KamMissWyl90}) that have not been included in this case. On the
other hand, $K$ denotes a particular such coupling that could
contribute at this order through the higher-dimensional operator
\be
   \frac{K}{F^2}[{\cal{O}}_1]_{\balpha\bbeta\bgamma\bdelta}
   \Tr\big[\chi^\dagger U+U^\dagger\chi\big].
\ee
Since this operator is of order $\chi_{cc}/F^2$ it must be included
according to our power counting rules. Thus we find that $c_2$ and
$d_2$ contain a logarithmic enhancement in $m_c$, while $c_1$ does
not.

A weak operator $[{\cal O}_2]_{\balpha\bbeta\bgamma\bdelta}$ of the
form shown in\,\eq{eq_O2_ChPT} can also contribute to the effective
weak Hamiltonian $\bar{\cal H}_{\rm w}^{\rm ChPT}$. For our choice of
mass matrix and the adopted power counting scheme, there is indeed a
tree-level effect of the form
\be
   -\frac{m_c^2}{F^2}
         \Big( \delta_{{\alpha}c}\delta_{{\delta}c}
               [\bar{\cal{O}}_3]_{\bbeta\bgamma} 
              +\delta_{{\beta}c}\delta_{{\gamma}c}
               [\bar{\cal{O}}_3]_{\balpha\bdelta} \Big),
\ee
which is of order $m_c^2$. In the region of small $m_c$ this is
parametrically smaller than the term with the same structure in
\eq{eq_O1_match}, which is multiplied by $d_2$. As the latter is
formally of order $m_c$, we drop the above contribution from now on.

We are now in a position to work out the effective weak Hamiltonian in
the $\threeby$ theory. To this end we start with the expression for
${\cal{H}}_{\rm w}^{\rm ChPT}$ in the SU(4)-symmetric theory,
\eq{eq_HwChPT}, drop the contribution from ${\cal{O}}_2$ for the
reason just mentioned, and write the remaining terms in the form
\be
  {\cal H}_{\rm w}^{\rm ChPT} =
  \frac{g^2_{\rm{w}}}{2M_W^2}(V_{us})^*V_{ud}
  \sum_{\sigma=\pm} g_1^\sigma c_{\alpha\beta\gamma\delta}
  (P_2^{\sigma}\,P_1^{\sigma})_{\alpha\beta\gamma\delta;\lambda\nu\rho\tau}
  [{\cal{O}}_1]_{\lambda\nu\rho\tau},
\label{eq_Hw_P1P2}
\ee
where the projectors $P_1^{\sigma}$ and $P_2^{\sigma}$ are defined in
eqs.~(\ref{eq_P1_def}) and~(\ref{eq_P2_def}). From the above
discussion we know that, to leading order in $m_c$, the matching is
achieved by replacing ${\cal{O}}_1$ by the right-hand side of
\eq{eq_O1_match}, with the coefficients $c_1,\,c_2$ and~$d_2$
specified as above. The expression for $\bar{\cal{H}}_{\rm w}^{\rm
ChPT}$ is obtained by inserting~\eq{eq_O1_match} into \eq{eq_Hw_P1P2}
and performing the decomposition into irreducible representations of
SU(3).\footnote{Our conventions for the SU(3) classification are
listed in Appendix~A of\,\cite{HerLai02}.} The resulting expression
for $\bar{\cal{H}}_{\rm{w}}^{\rm ChPT}$ is rather lengthy, and here we
simply quote the result for the physical choice of flavours,
characterised by $c_{suud}=1$ and $c_{sccd}=-1$:
\bea
   \bar{\cal{H}}_{\rm w}^{\rm ChPT} &=& 
   \frac{g_{\rm w}^2}{2M_W^2}(V_{us})^* V_{ud}\,\bigg\{
   c_1g_1^{+}[\widehat{\bar{\cal{O}}}_1^{+}]_{suud}
   +\left(\frac{c_1-5c_2}{5}g_1^{+}+(c_1-c_2)g_1^{-}\right)
   [\bar{\cal{R}}_1^{+}]_{sd}  \nonumber\\
   & & \phantom{\frac{g_{\rm w}^2}{2M_W^2}(V_{us})^* V_{ud}\,}
       -\frac{d_2}{2}(g_1^{+}
       +g_1^{-})[\widehat{\bar{\cal{O}}}_3]_{sd}\bigg\}.
\eea
In this expression the operators
$[\widehat{\bar{\cal{O}}}_1^{+}]_{suud}$ and
$[\bar{\cal{R}}_1^{+}]_{sd}$ are defined as
\bea
  [\widehat{\bar{\cal{O}}}_1^{+}]_{suud}
     \hspace{-0.2cm}&=&\hspace{-0.2cm}
     \half\left(
     [\bar{\cal{O}}_1]_{suud} +[\bar{\cal{O}}_1]_{sudu}
     -{\textstyle\frac{1}{5}} [\bar{\cal{O}}_1]_{s{\bkappa\bkappa}d}
     \right) ={\textstyle\frac{3}{5}}\left( [\bar{\cal{O}}_1]_{sudu}
     +{\textstyle\frac{2}{3}} [\bar{\cal{O}}_1]_{suud} \right),
     \label{eq_O1_bar_hat} \\
  {[\bar{\cal{R}}_1^{+}]_{sd}}
     \hspace{-0.2cm}&=&\hspace{-0.2cm}
     \half[\bar{\cal{O}}_1]_{s{\bkappa\bkappa}d}. 
\eea
The operator $[\widehat{\bar{\cal{O}}}_1^{+}]_{suud}$ transforms under
the irreducible representation of dimension~27 of SU(3), and in
\eq{eq_O1_bar_hat} we have listed two equivalent forms which are found
in the literature. The operators $[\bar{\cal{R}}_1^{+}]_{sd}$ and
$[\widehat{\bar{\cal{O}}}_3]_{sd}$ both transform under the
8-dimensional irreducible representation.

After inserting the expressions for the coefficients $c_1,\,c_2$ and
$d_2$ while keeping only logarithmic terms we finally arrive at
\bea
   \bar{\cal{H}}_{\rm w}^{\rm ChPT} &=&
   \frac{g_{\rm w}^2}{2M_W^2}(V_{us})^* V_{ud}\,\Bigg\{
   g_1^{+}[\widehat{\bar{\cal{O}}}_1^{+}]_{suud}  \nonumber\\
& & +\left[\frac{1}{5}g_1^{+}
   \bigg(1+\frac{15}{4}\frac{\chi_{cc}}{(4\pi F)^2}
                  \ln\frac{\Lambda_\chi^2}{\chi_{cc}}\bigg) +g_1^{-}
   \bigg(1+\frac{ 3}{4}\frac{\chi_{cc}}{(4\pi F)^2}
                  \ln\frac{\Lambda_\chi^2}{\chi_{cc}}\bigg)\right]
   [\bar{\cal{R}}_1^{+}]_{sd}  \nonumber\\
& &+\bigg[\frac{1}{4}(g_1^{+}+g_1^{-})\frac{\chi_{cc}}{(4\pi F)^2}
                  \ln\frac{\Lambda_\chi^2}{\chi_{cc}}\bigg]
   [\widehat{\bar{\cal{O}}}_3]_{sd}\Bigg\}.
\label{eq_Hwbar_expr}
\eea
This result illustrates the effects of the decoupling of the charm
quark.  One observes that the coefficients of the octet part (i.e. the
last two lines) contain an extra logarithmic factor,
$\chi_{cc}\ln\Lambda_\chi^2/\chi_{cc}$, which for $\chi_{cc} \ll
\Lambda_\chi^2$ is parametrically larger than the coefficient of the
27-plet (first line), even if the linear term of \eq{eq_c1_expr} were
kept in the latter. It is well known that the octet part only mediates
$\Delta{I}=1/2$ transitions, whereas the 27-plet contains both
$\Delta{I}=3/2$ and 1/2 contributions. Thus, the logarithmic terms in
front of the octet operators describe the departure of the
$\Delta{I}=1/2$ amplitude from the SU(4)-symmetric
situation.\footnote{Note, however, that the last term in
\eq{eq_Hwbar_expr} does not contribute to physical kaon
decays\,\cite{gammapm0,crewther}.}

We stress that these findings apply to the case of a moderately heavy
charm quark, which must be light enough for ChPT to be valid. Whether
or not the observed enhancement survives if the charm quark is tuned
towards its physical value must be studied in a lattice simulation.

%% file: sect7.tex
\section{Numerical results in the SU(4)-symmetric theory
  \label{sec_num}}

We now describe the findings of our numerical investigations in the
theory with a light, degenerate charm quark,
i.e. $m_u=m_d=m_s=m_c$. Thereby we will gain information as to what
extent physics at the intrinsic QCD scale of a few hundred MeV
contributes to the enhancement of the $\Delta{I}=1/2$ amplitude. The
effects of a heavier charm quark will be the subject of future
investigations.

In our numerical work we have used the quenched approximation, whose
deficits in this context have been discussed in
Sect.~\ref{sec_ChPT}. Although one may call into question any physical
interpretation of quenched numerical data, we regard the work
presented here primarily as a study to yield valuable information for
future simulations.

\subsection{General definitions}

Our task is the determination of the low-energy constants $g_1^{+}$
and $g_1^{-}$. Of particular interest is the ratio $g_1^{-}/g_1^{+}$,
which, according to \eq{eq_A0A2_ChPT}, determines the ratio of
amplitudes $A_0/A_2$ at leading order in ChPT. The determination of
LECs proceeds as usual, by computing suitable correlation functions in
simulations of lattice QCD and fitting the results to the expressions
obtained in ChPT. At non-vanishing lattice spacing such a procedure is
justified only if a fermionic discretisation is chosen which preserves
chiral symmetry. Here we have used the Neuberger-Dirac operator, which
was already discussed briefly in Sect.~\ref{sec_GWferms}.

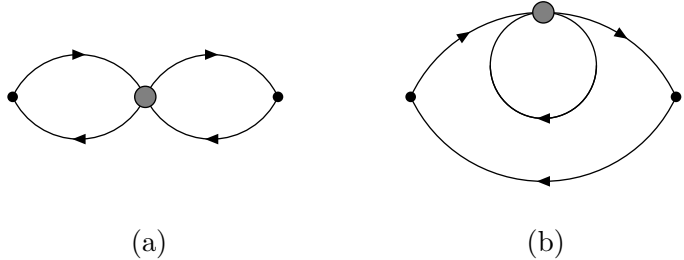
\begin{figure}
\begin{center}
\vspace{-1.4cm}
\begin{picture}(200,200)(0,0)
\Vertex(-30,120.00){2}
\Vertex(70,120.00){2}
\ArrowArcn(-5,108.34)(27.59,155,25)
\ArrowArcn(-5,131.66)(27.59,335,205)
\ArrowArcn(45,108.34)(27.59,155,25)
\ArrowArcn(45,131.66)(27.59,335,205)
\GCirc(20,120.00){4}{0.5}
\put(20,60){\makebox(3,10)[]{(a)}}

\Vertex(120,120.00){2}
\ArrowArcn(170,96.68)(55.18,90,25)
\ArrowArcn(170,96.68)(55.18,155,90)
\ArrowArcn(170,143.32)(55.18,335,205)
\BCirc(170,131.86){20}
\ArrowArcn(170,131.86)(20,360,180)
\GCirc(170,151.86){4}{0.5}
\Vertex(220,120.00){2}
\put(170,60){\makebox(3,10)[]{(b)}}

\end{picture}
\end{center}
\vspace{-2.4cm}
\caption{\footnotesize (a): the ``Figure-8'' diagram; (b): the ``Eye''
diagram. A grey circle denotes the insertion of ${\cal{Q}}_1^{\pm}$.
The left-handed current is inserted at the black circles.
\label{fig_diagrams}}
\end{figure}

From the previous sections of this paper it is clear that we are
particularly interested in correlation functions of left-handed
currents and four-quark operators. More specifically, we define the
non-singlet left-handed current as
\be
   [J_\mu(x)]_{\alpha\beta} = (\psibar_\alpha\gamma_\mu
                              P_-\psitilde_\beta)(x),
\ee
where $\alpha,\,\beta$ denote generic flavour indices, and the
modified quark field $\psitilde$ is given in \eq{eq_psitilde}. Our
earlier considerations show that we can concentrate on the operator
${\cal{Q}}_1^{\pm}$, whose definition we repeat here:\footnote{Note
that we have suppressed the superscript ``bare'' in comparison with
\eq{eq_Q1_bare}.}
\be
   {\cal Q}_1^{\pm} = \Big\{
   (\sbar\gamma_{\mu}P_{-}\tilde{u})(\ubar\gamma_{\mu}P_{-}\tilde{d})
\pm(\sbar\gamma_{\mu}P_{-}\tilde{d})(\ubar\gamma_{\mu}P_{-}\tilde{u})
   \Big\} - (u\,{\to}\,c).
\ee
We then consider the following two-point and three-point
correlators:\footnote{No summation over $\alpha, \beta$ is implied.}
\bea
   C(x_0) &=& \sum_{\xvec} \left\langle 
        [J_0(x)]_{\alpha\beta}[J_0(0)]_{\beta\alpha}
                           \right\rangle \label{eq_Cdef},   \\
   C_1^\pm(x_0,y_0) &=& \sum_{\xvec,\yvec}
   \left\langle [J_0(x)]_{du}\,[{\cal{Q}}_1^\pm(0)]\,[J_0(y)]_{us}
   \right\rangle.
\eea
After performing the Wick contractions, the correlators can be
expressed in terms of the quark propagator $S(x,y)$,
\be
   \Sprop{\alpha}{x}{y} \equiv \left\langle
   \psitilde_\alpha(x)\psibar_\alpha(y)\right\rangle
   =\left\{(1-\half\abar D)D_{m_\alpha}^{-1}\right\}(x,y),
\ee
where
\be
   D_{m_\alpha}= (1-\half\abar m_\alpha)D+m_\alpha
\ee
is the massive Neuberger-Dirac operator with bare mass $m_\alpha$. The
left-handed propagator is then given by~\cite{numeps}
\be
   P_-\Sprop{\alpha}{x}{y}P_+ =(1-\half\abar m_\alpha)^{-1} \left\{
        P_-(\gamma_5 D_{m_\alpha})^{-1}P_+ \right\}(x,y).
\ee
Since here we restrict ourselves to studying the case where all quark
masses are degenerate, we will omit from now on the flavour label on
propagators. The expression for the two-point function $C(x_0)$ then
reads
\be
   C(x_0) = \sum_{\xvec}\left\langle 
     \Tr \left\{\gamma_0 P_- S(x,0)^\dagger P_+
                \gamma_0 P_- S(x,0) P_+ \right\}
                              \right\rangle.
\label{eq_ctwo}
\ee
Turning now to the three-point correlator, we show in
Fig.~\ref{fig_diagrams} the two types of diagrams that are obtained
after performing the Wick contractions. An important consequence of
restricting to the SU(4)-symmetric case is the fact that contributions
from the so-called ``Eye''-diagram vanish identically. Thus, only
diagrams of the ``Figure-8'' type must be considered, and the
expression for $C_1^{\pm}$ in terms of quark propagators becomes
\bea
    C_1^\pm(x_0,y_0) &=&\sum_{\xvec,\yvec}\sum_{\mu=0}^3\Big\{
    \Big\langle 
    \Tr\{\gamma_\mu P_{-}S(x,0)^\dagger \gamma_0 P_{-}S(x,0)\} 
    \Tr\{\gamma_\mu P_{-}S(y,0)^\dagger \gamma_0 P_{-}S(y,0)\} 
    \Big\rangle  \nonumber \\
    & & \phantom{\sum}
 \mp\Big\langle
    \Tr\{\gamma_\mu P_{-}S(x,0)^\dagger \gamma_0 P_{-}S(x,0)
        \gamma_\mu P_{-}S(y,0)^\dagger \gamma_0 P_{-}S(y,0)\}
    \Big\rangle\Big\}.  \label{eq_C1}
\eea
Thus, correlation functions of operators which transform under
irreducible representations of dimensions~84 and~20 are obtained by
taking appropriate linear combinations of colour-connected and
colour-disconnected contractions. As an aside we remark that the
correlator $C_1^{+}(x_0,y_0)$ can also be used to compute the
$B$-parameter $B_K$.

In accordance with the relation between ${\cal{Q}}_1^\pm$ and
${\cal{O}}_1$ (see \eq{eq_Q1_O1rel}), we note that the combination
$k_1^\pm C_1^\pm(x_0,y_0)$ corresponds to $2g_1^\pm\hat{\cal
C}_1^\pm(x_0,y_0)$ in the low-energy description, while $k_1^\pm
C_1^\pm(x_0,y_0)/[C(x_0)C(y_0)]$ corresponds to $g_1^\pm
K^\pm(x_0,y_0)$ (see eqs.~(\ref{eq_C1pm_ChPT}) and
(\ref{eq_Kpm_def})).

\subsection{Technical details}

In order to evaluate the correlation functions defined above, we have
computed left-handed quark propagators on quenched background gauge
configurations, using the Neuberger-Dirac operator with~$s=0.4$ (see
\eq{eq_Dneu}). We have employed the numerical techniques described in
ref.~\cite{numeps}, including the approximation of the inverse square
root in $D$ by a minmax polynomial, the determination of the index
$\nu$ of a given gauge configuration via the counting of zero modes,
and the computation of a number of low modes of $P_{\sigma}D^\dagger
DP_{\sigma}$, where $P_{\sigma}$ projects onto the chirality sector
without zero modes. The calculation of the left-handed propagator was
accelerated using low-mode preconditioning as described in
\cite{numeps}. Furthermore, the calculation was arranged such that the
necessary Conjugate Gradient inversions were always performed in the
chirality sector without zero modes.

In several recent publications \cite{NIC_corr,lma} it was reported
that correlation functions of left-handed currents show strong
statistical fluctuations if the quark mass becomes of order
$({\Sigma}V)^{-1}$ or smaller. The origin of these fluctuations could
be traced to the low modes of the Dirac operator. For instance, if
$m\simeq({\Sigma}V)^{-1}$ the contributions of a few eigenmodes to a
given observables can be substantial, and the intrinsic space-time
fluctuations of the eigenmodes then induce a large variance in the
Monte Carlo estimate. In ref.~\cite{lma} we proposed an exact
technique which is able to reduce these fluctuations significantly,
and which involves taking volume averages of the contributions of a
certain number of low modes, $n_{\rm low}$, to the respective
correlators. This technique,\footnote{Similar methods were discussed
in refs.~\cite{DeG_Sch04} and~\cite{Edw_lat01}.} dubbed ``low-mode
averaging'', allows for the computation of two-point functions with
controlled statistical errors in all topological sectors if $n_{\rm
low}\;\lesssim\;7$, for quark masses around $m\sim(\Sigma V)^{-1}$. We
note, though, that our technique cannot cure the fluctuations caused
by extremely small eigenvalues of $D$, which are expected to occur
with a non-negligible probability if $m\ll(\Sigma V)^{-1}$. As already
remarked in~\cite{lma}, the solution to this problem may require the
incorporation of the low-mode contribution to a particular observable
into the importance sampling process.

For this study we extended the technique of low-mode averaging to the
three-point functions corresponding to the diagrams in
Fig.\,\ref{fig_diagrams}. It turned out, however, that low-mode
averaging for three point functions requires a more careful tuning of
its free parameters, in order to observe a significant reduction of
statistical errors. Preliminary results of a detailed study have shown
that masses $m\leq (\Sigma V)^{-1}$ can be reached if the number of
low modes treated exactly is increased by a factor $2-3$ compared to
the values in ref.\,\cite{lma}.\footnote{We thank C. Pena and
J. Wennekers for their contribution on this point.} For the purpose
of this paper we have, instead, focused our attention on moderately
light quark masses, corresponding to the $p$-regime. The parameters of
our simulations are listed in Table\,\ref{tab_params}.

\begin{table}[t]
\begin{center}
\begin{tabular}{c l c c c c r}
\hline
Lattice & $\beta$ & $L/a$ & $T/a$ & $n_{\rm{low}}$  & $L[\fm]$
        & configs.  \\
\hline
A & 6.0    & 12 & 40 &      7   & 1.12 & 751 \\
B & 5.8485 & 12 & 30 &      5   & 1.49 & 638 \\
\hline
\end{tabular}
\caption{\footnotesize Simulation parameters, including the spatial
  and temporal lattice sizes, the number of low modes of the Dirac
  operator which are treated exactly, and the number of gauge
  configurations. \label{tab_params}}
\end{center}
\end{table}

\begin{table}
\begin{center}
\begin{tabular}{c c c c r@{.}l}
\hline
Lattice & $am$ & $a\mps$ & $a\Fp$
 & \multicolumn{2}{c}{$C_1^{-}/C_1^{+}$} \\
\hline
A & 0.030 & 0.2248(30) & 0.03434(37) & 2&46(40) \\
  & 0.040 & 0.2522(28) & 0.03527(36) & 1&94(25) \\
  & 0.050 & 0.2772(27) & 0.03623(36) & 1&67(18) \\
  & 0.060 & 0.3005(26) & 0.03719(37) & 1&52(15) \\
  & 0.070 & 0.3224(27) & 0.03815(39) & 1&41(12) \\
\hline
B & 0.040 & 0.2628(28) & 0.04149(45) & 2&03(25) \\
  & 0.053 & 0.2964(28) & 0.04244(46) & 1&75(17) \\
  & 0.066 & 0.3268(28) & 0.04339(47) & 1&57(12) \\
  & 0.078 & 0.3529(27) & 0.04426(48) & 1&45(10) \\
  & 0.092 & 0.3816(27) & 0.04527(49) & 1&35(8)  \\
\hline
\end{tabular}
\caption{\footnotesize Results for pseudoscalar meson masses and the
  ratio of three-point functions $C_1^{-}/C_1^{+}$ at several values
  of the bare quark mass. \label{tab_results}}
\end{center}
\end{table}

\subsection{Results}

In order to determine pseudoscalar masses and decay constants we
fitted the two-point function $C(x_0)$ to the expression \cite{lma}
\be
   C(x_0) = \half\mps\Fp^2\,
   \frac{\cosh\big[(T/2-|x_0|)\mps\big]}{2\sinh[T\mps/2]}.
\ee
For lattice~A these fits were performed for timeslices in the range $7
\leq x_0/a \leq 19$, while for lattice~B we used $7\leq x_0/a \leq
14$. The results are listed in Table~\ref{tab_results}. We note that
the values obtained on lattice~A for bare masses $am=0.04$ and~0.06
can be directly compared with ref.~\cite{lma}, where a larger spatial
volume corresponding to $L=1.49\,\fm$ was used. The results for the
pseudoscalar decay constant agree within errors, while the
pseudoscalar meson mass at $am=0.040$ is larger by 4\% (1.8$\sigma$)
on the smaller spatial volume, which indicates a small finite-volume
effect in the pseudoscalar correlator.

\begin{figure}
\begin{center}
\psfig{file=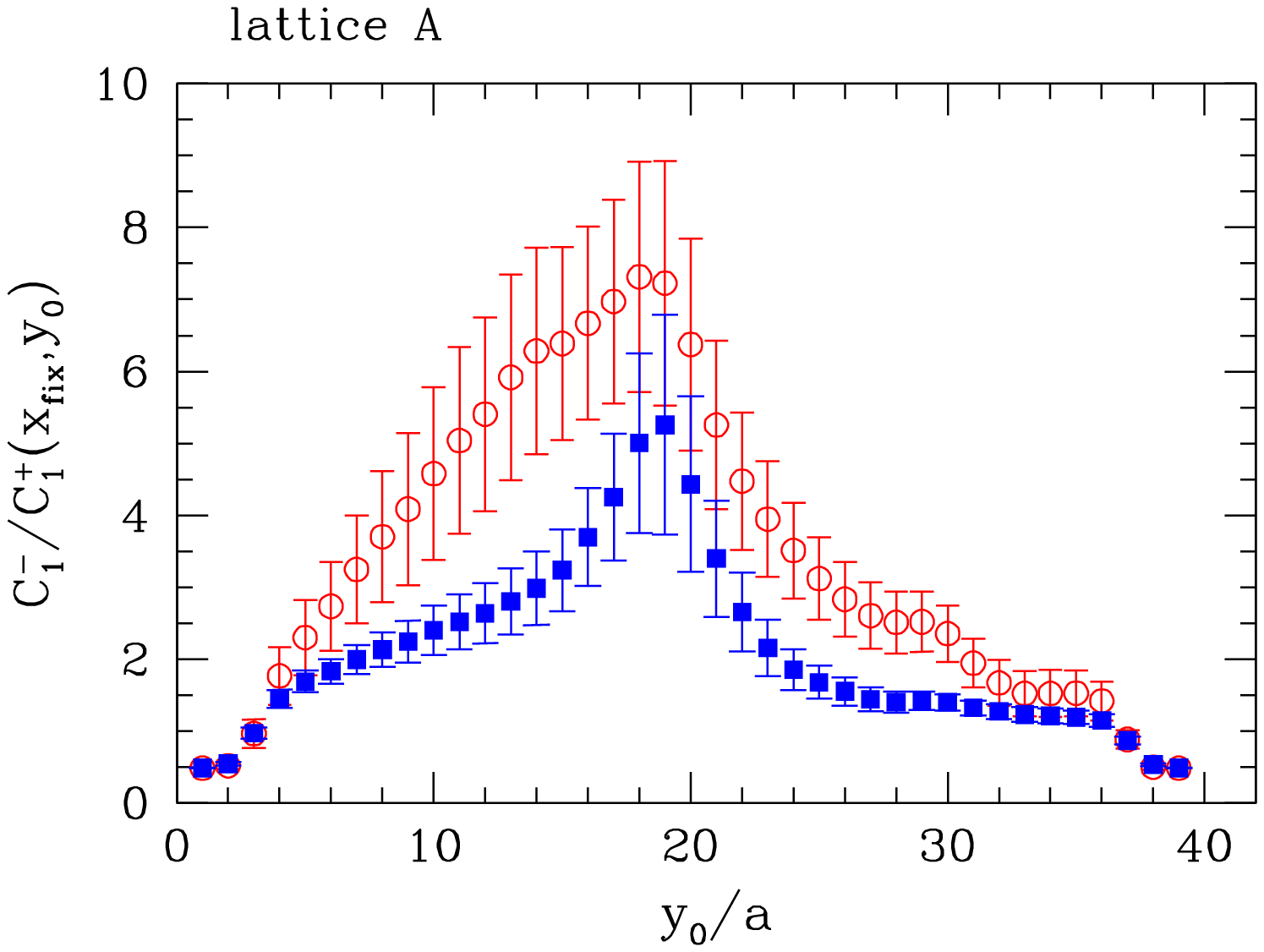,width=7.cm}
\psfig{file=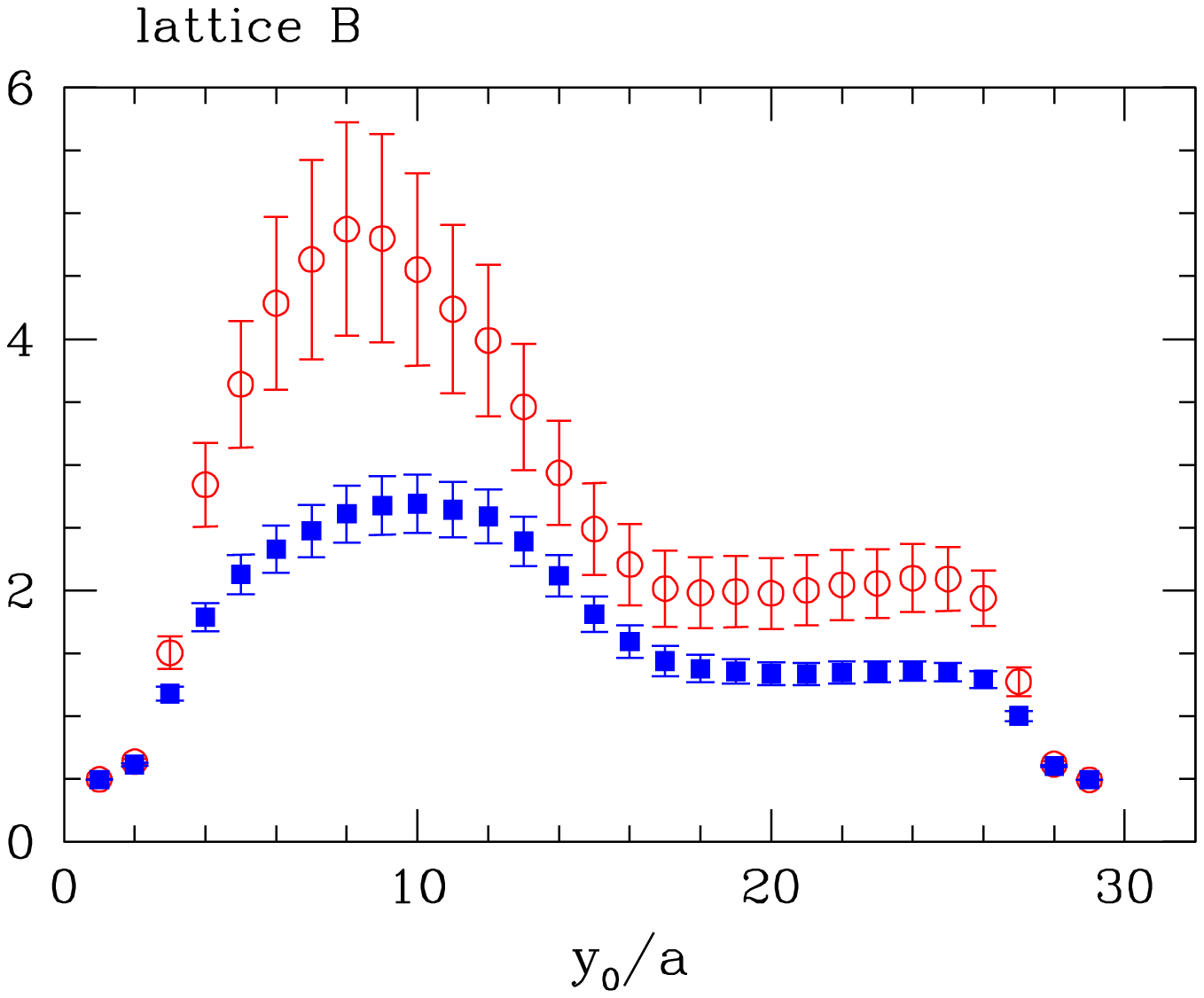,width=7.cm}
\caption{\footnotesize The ratio $C_1^{-}(x_{\rm
fix},y_0)/C_1^{+}(x_{\rm fix},y_0)$ as a function of $y_0/a$ for
lattices~A (left panel) and~B (right panel). Open circles and
full squares correspond to the smallest and biggest
quark mass, respectively. \label{fig_3pt_sig}}
\end{center}
\end{figure}

In order to isolate the asymptotic behaviour of the ratio of
three-point functions, $C_1^{-}(x_0,y_0)/C_1^{+}(x_0,y_0)$, we first
fixed $x_0$ at $x_0=x_{\rm fix}=T/4$ (actually, $x_{\rm fix}=8a$ for
lattice~B) and looked for a plateau in $y_0$ around
$T-x_{\rm{fix}}$. Figure~\ref{fig_3pt_sig} shows the quality of the
plateau for both runs at two values of the bare quark mass. By fitting
the ratio $C_1^{-}/C_1^{+}$ to a constant for $28 \leq y_0/a \leq 30$
(lattice~A) and $18 \leq y_0/a \leq 25$ (lattice~B), respectively, we
obtain the results listed in the last column of
Table~\ref{tab_results}.

Our results show that there is a clear numerical signal for the ratio
of correlators proportional to $g_1^{-}/g_1^{+}$ in the $\fourby$
symmetric theory. The typical statistical error in the studied range of
quark masses is at the level of 10\%, which should be sufficient for
the purpose of establishing whether physics at the intrinsic QCD scale
makes a significant contribution to the $\Delta{I}=1/2$ rule.

\begin{figure}[t]
\begin{center}
\psfig{file=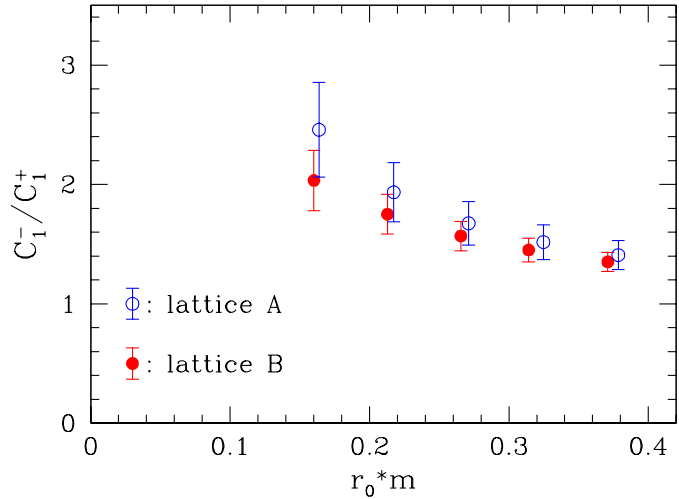,width=9.cm}
\caption{\footnotesize The fitted value of the ratio $C_1^{-}/C_1^{+}$
  plotted versus the bare quark mass in units of $r_0$. The symbols
  are slightly displaced for clarity. \label{fig_20over84}}
\end{center}
\end{figure}


In Fig.~\ref{fig_20over84} we plot the results for the asymptotic
value of $C_1^{-}/C_1^{+}$ for both our lattices as a function of the
bare quark mass in units of the hadronic radius $r_0$
\cite{r0_refs}. At the smallest quark mass the ratio of correlators of
operators transforming under irreducible representations of
dimensions~20 and~84 is clearly greater than one. Furthermore, one
observes a clear trend for this ratio to increase as the chiral limit
is approached.

At leading order in ChPT the connection between $C_1^{-}/C_1^{+}$ and
the ratio $g_1^{-}/g_1^{+}$ of LECs is given by
\be
   \frac{g_1^{-}}{g_1^{+}} = \frac{k_1^{-}}{k_1^{+}}
   \left.\frac{C_1^{-}(x_0,y_0)}{C_1^{+}(x_0,y_0)}\right|_{\rm ren},
\label{eq_20over84_LO}
\ee
where the subscript ``ren'' reminds us that the four-quark operators
in the lattice-regularised theory must be related to a particular
continuum scheme. Thus, in order to estimate $g_1^{-}/g_1^{+}$ it
suffices to multiply $C_1^{-}/C_1^{+}$ by the relevant short-distance
factors. When interpreted in this way, our results indicate a
significant enhancement of the $\Delta{I}=1/2$ amplitude from
long-distance QCD effects alone. However, the prediction of
leading-order ChPT in \eq{eq_20over84_LO} implies that there should be
no significant mass dependence in the ratio, a behaviour which is not
reflected in our results.

Clearly, NLO effects must be taken into account. In the $p$-regime
they depend on many unknown effective couplings, while, as we have
shown in this paper, they can be computed in the $\epsilon$-regime
without adding any extra parameters (see eqs.~(\ref{eq_g1pm_R2})
and~(\ref{eq_Hexp})). Since we lack any data in the $\epsilon$-regime,
the NLO matching of $C_1^{-}/C_1^{+}$ to $g_1^{-}/g_1^{+}$ cannot be
performed here. However, in order to illustrate how the NLO formulae
might be used, we note that for lattice~B the correction factor
$H(x_0,y_0)$ evaluates to~2.262, independent of $x_0$ and $y_0$. Thus,
if long-distance QCD effects alone were to enhance $g_1^{-}/g_1^{+}$
significantly, say, by a factor~2, then the ratio $C_1^{-}/C_1^{+}$
would have to be as large as $4-5$ in the $\epsilon$-regime.
 
We note, though, that the geometrical factor $\rho\equiv T/L$ enters
$H(x_0,y_0)$, such that for space-time geometries with $T\gg L$ the
NLO corrections can become large. For instance, on lattice~B we have
$\rho=2.5$, and one might expect that higher orders in the
$\epsilon$-expansion cannot be neglected in this case
(cf. Figs.\,\ref{fig_ChPT_3pt} and \ref{fig_3pt_rat}).

Therefore, while we observe a good numerical signal for the ratio
$C_1^{-}/C_1^{+}$, a quantitative interpretation of our results is not
yet possible, since our masses are not in a regime where computable
NLO corrections can be applied. Future simulations should therefore
concentrate on the determination of $C_1^{-}/C_1^{+}$ for smaller
quark masses.  Moreover, choosing a small value of $T/L$, while
keeping the spatial box length reasonably large
(e.g. $L\;\gtaeq\;1.5\,\fm$) will lead to better convergence
properties of ChPT in the $\epsilon$-regime.

%% file: sect8.tex
\section{Summary and conclusions \label{sec_concl}}

We have outlined a computational strategy which allows to disentangle
and quantify several possible origins of the $\Delta{I}=1/2$ rule
using lattice simulations of QCD. The main idea is to consider the
$\Delta{S}=1$ weak interactions with an active charm quark and to
compute correlation functions for $K\to\pi$ transitions, whose values
are subsequently matched to the LECs which parameterise the weak
Hamiltonian in the effective low-energy theory. While the
mass-degenerate case, $m_u=m_d=m_s=m_c$, allows to quantify
contributions to the $\Delta{I}=1/2$ enhancement from long-distance
effects that are not due to a large charm quark mass, the effects of
physics at scales around $m_c$ can be estimated by increasing $m_c$
above $m_u$ and evaluating the ``Eye''-diagram in addition to the
``Figure-8''-graph. From the behaviour of the ratio of amplitudes as a
function of $m_c$ one will be able to infer whether QCD long-distance
effects are merely amplified by the splitting between up and charm
quark masses, or whether the $\Delta{I}=1/2$ rule is due to modes at a
specific energy scale, which should manifest itself as a sharp change
in magnitude.

A key ingredient in our strategy is the use of fermionic
discretisations which preserve chiral symmetry at non-zero lattice
spacing (Ginsparg-Wilson fermions). In this case the renormalisation
patterns of the relevant dimension-6 operators which are found in the
continuum carry over to the lattice theory without
modification. Moreover, the matching to the expressions of ChPT can be
performed in a conceptually clean manner. Furthermore, the
$\epsilon$-regime of ChPT offers a firm theoretical framework for the
determination of LECs: owing to the specific chiral counting rules of
the $\epsilon$-expansion, no additional interaction terms are
generated in the effective weak part at NLO. By contrast, for the
conventional chiral expansion ($p$-expansion) the matching to QCD at
NLO involves coupling terms whose coefficients are unknown.

Our studies have revealed, though, that fairly large volumes are
required in order to guarantee reliable results. The slow convergence
of ChPT in the $\epsilon$-regime is indicated in
Figs. \ref{fig_ChPT_3pt} and \ref{fig_3pt_rat}, which demonstrate that
corrections are relatively small only for $L\;\gtaeq\;2.0\,\fm$.  The
use of asymmetric lattices slows down the convergence further. Since
lattice simulations are typically performed with $T>L$, in order to be
able to isolate the asymptotic behaviour of Euclidean correlation
functions, whilst keeping the total number of lattice sites at a
manageable level, this may render future numerical simulations
relatively expensive.

We have performed numerical simulations in the SU(4)-symmetric case in
the quenched approximation, with masses corresponding to the
$p$-regime. The results show that the matching of $C_1^{-}/C_1^{+}$ to
$g_1^{-}/g_1^{+}$ at leading order is clearly insufficient. On the
other hand, for the reasons explained above, adjusting the kinematical
variables so that NLO matching can be performed reliably in the
$\epsilon$-regime is numerically expensive. In fact, the computational
cost of simulating the kinematical regime where the $p$-expansion at
leading order is valid might be even of comparable
size.\footnote{Various strategies for matching simultaneously for the
LO and for a number of NLO couplings in the conventional SU(3)-case
have been outlined in \cite{NLO_SU3}.}  Which of these alternatives is
to be preferred must be decided by future simulations.

We have also investigated the effects of a heavier charm quark in
ChPT. These studies have shown that there is indeed a larger
contribution to $\Delta{I}=1/2$ transitions when the charm quark mass
is increased above the masses of the $u,\,d$ and $s$ quarks but kept
below the chiral scale $\Lambda_\chi$. Since only moderately heavy
charm quark masses can be studied safely in ChPT, numerical
simulations must be performed to confirm this result. This will be the
subject of our future simulations, in which we will evaluate the
necessary ``Eye''-diagrams. Furthermore, we intend to improve on the
systematics of our earlier runs, by using lattices with smaller $T/L$
and going closer to the chiral limit.

\subsection*{Acknowledgements}
We are indebted to Martin L\"uscher, whose participation in the early
stages of this project was instrumental to our study. We would like to
thank him for many illuminating discussions and a careful reading of
the manuscript. We are also grateful to Christian Hoelbling, Karl
Jansen and Laurent Lellouch for interesting discussions. Our
simulations were performed on PC clusters at DESY Hamburg, the
Leibniz-Rechenzentrum der Bayerischen Akademie der Wissenschaften, the
Max-Planck-Institut f\"ur Plasmaphysik in Garching and at the
University of Valencia. We wish to thank all these institutions for
support and the staff of their computer centers for technical
help. L.G. acknowledges partial support by the EU under contract
HPRN-CT-2002-00311 (EURIDICE). P.H. was supported by the CICYT
(Project No. FPA2002-00162) and by the Generalitat Valenciana (Project
No. CTIDIA/2002/5).

%% file: appendix1.tex
\section{Four-quark representations\label{app1}}

In this appendix, we provide for completeness some aspects of the
SU(4) classification of four-quark operators. For further details see,
for instance, the complete presentation in \cite{notes:SU4rep}. We
consider an operator $O_{\alpha\beta\gamma\delta}$ which transforms
under the $\bf 4^*\otimes4^*\otimes4\otimes4$ representation of SU(4)
and decompose it into its irreducible parts. To this end we define the
projected operators
\bea
& & O_{\alpha\beta\gamma\delta}^\sigma\equiv
   (P_1^\sigma)_{\alpha\beta\gamma\delta;\lambda\nu\rho\tau}
    O_{\lambda\nu\rho\tau},  \label{eq_P1}\\
& & \widehat{O}_{\alpha\beta\gamma\delta}^\sigma\equiv
   (P_2^\sigma)_{\alpha\beta\gamma\delta;\lambda\nu\rho\tau}
    O_{\lambda\nu\rho\tau}^\sigma, \label{eq_P2}
\eea
where $\sigma=\pm1$. The projectors $P_1^\sigma, P_2^\sigma$ are given
by
\bea
 (P_1^\sigma)_{\alpha\beta\gamma\delta;\lambda\nu\rho\tau} &\equiv&
\quarter(\delta_{\alpha\lambda}\delta_{\beta\nu}
  +\sigma\delta_{\alpha\nu}\delta_{\beta\lambda})
        (\delta_{\gamma\rho}\delta_{\delta\tau}
  +\sigma\delta_{\gamma\tau}\delta_{\delta\rho}), \label{eq_P1_def} \\
 (P_2^\sigma)_{\alpha\beta\gamma\delta;\lambda\nu\rho\tau} &\equiv&
    \delta_{\alpha\lambda}\delta_{\beta\nu}\delta_{\gamma\rho}
    \delta_{\delta\tau} +\frac{1}{(4+2\sigma)(4+\sigma)}
    (\delta_{\alpha\gamma}\delta_{\beta\delta}  +\sigma
     \delta_{\alpha\delta}\delta_{\beta\gamma})
    \delta_{\lambda\rho}\delta_{\nu\tau}\phantom{\frac{1}{4+2\sigma}}
    \nonumber
\eea
\be
 -\frac{1}{4+2\sigma}
    (\delta_{\alpha\gamma}\delta_{\beta\nu}\delta_{\delta\tau}
     \delta_{\lambda\rho}
    +\delta_{\beta\delta}\delta_{\alpha\lambda}\delta_{\gamma\rho}
     \delta_{\nu\tau} +\sigma
     \delta_{\alpha\delta}\delta_{\beta\nu}\delta_{\gamma\tau}
     \delta_{\lambda\rho} +\sigma
     \delta_{\beta\gamma}\delta_{\alpha\lambda}\delta_{\delta\rho}
     \delta_{\nu\tau}).  \label{eq_P2_def}
\ee
Furthermore we define
\bea
& & S^\sigma = O_{\kappa\lambda\kappa\lambda}^\sigma, \\
& & R^\sigma_{\alpha\gamma} = O_{\alpha\lambda\gamma\lambda}^\sigma
    -\quarter\delta_{\alpha\gamma}S^\sigma.
\eea
Any four-quark operator $O_{\alpha\beta\gamma\delta}$ can then be
decomposed into irreducible representations according to
\bea
  O_{\alpha\beta\gamma\delta}&=&\sum_{\sigma=\pm}\left\{
  \widehat{O}_{\alpha\beta\gamma\delta}^\sigma+
  \frac{1}{4(4+\sigma)}(\delta_{\alpha\gamma}\delta_{\beta\delta}
                 +\sigma\delta_{\alpha\delta}\delta_{\beta\gamma})
  S^\sigma\right. \nonumber\\
& &  \phantom{\sum_{\sigma=\pm}\Big\{
  \widehat{O}_{\alpha\beta\gamma\delta}^\sigma}
  \left.
  +\frac{1}{4+2\sigma}(\delta_{\alpha\gamma}R^\sigma_{\beta\delta}
                      +\delta_{\beta\delta}R^\sigma_{\alpha\gamma}
                +\sigma\delta_{\alpha\delta}R^\sigma_{\beta\gamma}
                +\sigma\delta_{\beta\gamma}R^\sigma_{\alpha\delta})
  \right\}.\phantom{\alpha}
\eea
The operators $\widehat{O}_{\alpha\beta\gamma\delta}^{\pm}$,
$R^{\pm}_{\alpha\gamma}$ and $S^\pm$ transform according to
representations of the following dimensions:
\be
   \begin{array}{lcr lcr}
   \widehat{O}_{\alpha\beta\gamma\delta}^{+} &\hbox{:}& {\bf 84}, &
   \quad\qquad R^\pm_{\alpha\gamma} &\hbox{:}& {\bf 15}, \\
   \widehat{O}_{\alpha\beta\gamma\delta}^{-} &\hbox{:}& {\bf 20}, &
   \quad\qquad S^\pm                &\hbox{:}& {\bf 1}.
   \end{array}
\ee
In \eq{eq_Q1pm}, the operators ${\cal Q}_1^\pm$ have been defined for
the relevant physical flavour assignments. Their relations to the
projected operators $\widehat{O}_1^\pm$ are given by
\be
   {\cal Q}_1^\pm =
   2([\widehat{O}_1^\pm]_{suud}-[\widehat{O}_1^\pm]_{sccd}),
\label{eq_Q1_O1rel}
\ee
where $\widehat{O}_1^\pm$ are obtained from the generic four-quark
operator $[O_1]_{\alpha\beta\gamma\delta}$ via eqs.~(\ref{eq_P1})
and~(\ref{eq_P2}).

%% file: appendix2.tex
\section{Exact chiral symmetry and CP \label{app_CP}}

In this appendix we report the argument spelled out in
\cite{notes:4quark}, which shows that in lattice QCD with
Ginsparg-Wilson fermions, the CP symmetry remains a useful tool in the
analysis of operator mixing, if a chiral operator basis is
chosen. Assuming that $D$ satisfies the Ginsparg-Wilson relation, we
note that the quark action
\be
   S_{\rm F}=a^4\sum_x \left\{ \psibar(x)[D\psi](x) 
   +\psibar(x)\big(MP_{+}+M^{\dagger}P_{-}\big) \big[
    \big(1-{\textstyle\frac{1}{2}}{\abar}D\big)\psi\big](x) \right\}
\ee
is invariant under the exact $\rm SU(4)_L\times SU(4)_R$ symmetry
group, provided the quark mass matrix $M$ is transformed according to
the $\bf 4\otimes4^*$ representation. If the mass matrix $M$ is chosen
real and diagonal, the action is also invariant under parity and
charge conjugation. The restriction to this case is not important,
since chiral transformations can always be used to diagonalise
$M$. However, four-quark operators such as
\be
   {O}^\pm_{\alpha\beta\gamma\delta}=
   \big(\psibar_\alpha\gamma_\mu P_{-}\psitilde_\gamma\big)
   \big(\psibar_\beta\gamma_\mu  P_{-}\psitilde_\delta\big)
   \pm (\gamma\leftrightarrow\delta)
\ee
do not transform in a simple way under CP. At the level of correlation
functions of local fields, this is not a problem, because one can
substitute
\be
   \psitilde(x)=\big(1-{\textstyle\frac{1}{2}}{\abar}M\big)^{-1}
   \left\{\psi(x)-{\textstyle\frac{1}{2}}{\abar}
          \frac{{\delta}S_{\rm F}}{\delta\psibar(x)}\right\}
\ee
and perform a partial integration in the functional integral to
eliminate the second term on the right-hand side. One then obtains the
same correlation functions up to contact terms, but with the modified
field $\psitilde$ replaced by
$(1-{\textstyle\frac{1}{2}}{\abar}M)^{-1}\psi$. 

In the case of the operators ${O}^\pm_{\alpha\beta\gamma\delta}$
one can show that all contact terms vanish after contracting the Dirac
indices. When inserted in correlation functions of local fields at
non-zero distances, one obtains the operator identity
\bea
   {O}^\pm_{\alpha\beta\gamma\delta} &=& 
   \big(1-{\textstyle\frac{1}{2}}{\abar}m_\gamma\big)^{-1}
   \big(1-{\textstyle\frac{1}{2}}{\abar}m_\delta\big)^{-1} \nonumber\\
   & &\times\Big\{
   \big(\psibar_\alpha\gamma_\mu P_{-}\psi_\gamma\big)
   \big(\psibar_\beta\gamma_\mu  P_{-}\psi_\delta\big)
   \pm(\gamma\leftrightarrow\delta) \Big\}.
\eea
Applying a CP transformation yields
\be
   {O}^\pm_{\alpha\beta\gamma\delta}(x)
   \stackrel{\hbox{CP}}{\longrightarrow}
   \frac{(1-{\textstyle\frac{1}{2}}{\abar}m_\alpha)
         (1-{\textstyle\frac{1}{2}}{\abar}m_\beta)}
        {(1-{\textstyle\frac{1}{2}}{\abar}m_\gamma)
         (1-{\textstyle\frac{1}{2}}{\abar}m_\delta)}
   \,{O}^\pm_{\gamma\delta\alpha\beta}(\tilde{x}),
\ee
where $\tilde{x}=(x_0,-x_1,-x_2,-x_3)$, and it is understood that the
rule only applies in local correlation functions. The argumentation in
the case of two-quark operators proceeds along exactly the same lines.

%% file: appendix3.tex
\section{Perturbative results for the anomalous dimensions 
$\gamma^{\sigma}(g)$\label{app3}}

For convenience we recall that the RG $\beta$-function has a
perturbative expansion
\be
\beta(g)=-g^3\sum_{k=0}^{\infty}b_k g^{2k}\,,
\ee
with the universal one-- and two--loop coefficients:
\bea
b_0&=&\frac{1}{(4\pi)^2}\left(\frac{11}{3}\Nc-\frac23\Nf\right)\,,
\\
&&\nonumber
\\
b_1&=&\frac{1}{(4\pi)^4}\left(\frac{34}{3}\Nc^2
-\Nf\left\{\frac{13}{3}\Nc-\frac{1}{\Nc}\right\}\right)\,.
\label{betauniv} 
\eea
The anomalous dimension of ${\cal{Q}}_1^\sigma$ has a perturbative
expansion of the form 
\be
\gamma^{\sigma}(g)=g^2\sum_{k=0}^\infty \gamma^{\sigma}_k g^{2k}.
\label{gamma}
\ee
The one--loop coefficients $\gamma^{\sigma}_0$ were 
first obtained by Gaillard and Lee \cite{gammapm0},
\be
\gamma^{\sigma}_0=-\frac{1}{(4\pi)^2}\frac{6\sigma(\Nc-\sigma)}{\Nc}\,.
\label{gailee}
\ee
The two--loop coefficients depend on the scheme and on the way
$\gamma_5$ is handled. The first such two--loop computation was
performed by Altarelli, Curci, Martinelli and Petrarca
\cite{Alta_etal} using dimensional reduction \cite{Siegel}. The
computation in the HV(MS) scheme was carried out much later with the
result \cite{BurasWeisz90}:
\be
\gamma^{\sigma}_1=\frac{1}{(4\pi)^4}
\frac{(\Nc-\sigma)}{2\Nc}\left[-\frac{88}{3}\Nc^2+21+\frac{16}{3}\Nc\Nf
-\sigma\left\{\frac{157}{3}\Nc+\frac{57}{\Nc}
-\frac{28}{3}\Nf\right\}\right]\,.
\label{gamma1hv}
\ee 
Denoting the operators in the HV scheme by $({\cal{Q}}_1^\sigma)_{\rm
HV}$, the renormalised operators of any other satisfactory scheme S
should be related to these merely by a finite multiplicative
renormalisation,
\be
 ({\cal{Q}}_1^\sigma)_{\rm S}={\cal X}^\sigma_{\rm S}(g)
 ({\cal{Q}}_1^\sigma)_{\rm HV}\,.
\ee
Assuming that the normalisations are such that ${\cal X}^\sigma_{\rm
S}=1$ at tree level, the coefficients have a perturbative expansion of
the form
\be
 {\cal X}^\sigma_{\rm S}(g)=1+\sum_{k=1}^\infty x^\sigma_{\rm
 S;k}g^{2k}\,. 
\ee
The one-loop coefficients have been computed in various schemes,
e.g. in the RI scheme\footnote{This is defined by demanding that the
renormalised 4--point vertex function in the Landau gauge at equal
external momenta $p$ with $p^2=\mu^2$ (where $\mu$ is usually set
equal to the standard renormalisation scale in the $\msbar$ scheme) be
equal to the tree level function.}  one has
\be
x^\sigma_{\rm RI;1}=-\frac{1}{(4\pi)^2}\frac{(\Nc-\sigma)}{2\Nc}
\left[4\Nc-6\sigma\left\{1-4\ln(2)\right\}\right]\,.
\label{xt1ri}
\ee
For Ginsparg-Wilson fermions, as defined in Sect.~\ref{sec_Hw_latt},
one has
\be
x^\sigma_{\rm GW;1}=\frac{1}{(4\pi)^2}\frac{(\Nc-\sigma)}{2\Nc}
\left[X_0\sigma+X_1 \Nc\right]\,,
\label{xt1latt}
\ee
with
\bea
    X_0&=&4B_{\rm V}-2B_{\rm S}+2B_\psi+6\,, \\
    X_1&=&2B_{\rm V}+2B_\psi-4\,,
\eea
where the functions $B_{\rm V},B_{\rm S},B_\psi$ depend on the
parameter $s$, and some numerical values can be found in Table~1 of
ref.~\cite{SteLeo00}. Corresponding values for the $X_i$ are given in
Table~\ref{tabx}.

The anomalous dimensions of the operators in scheme S are
related to those in the HV scheme by
\be
\gamma^\sigma_{\rm S}=\gamma^\sigma+\beta(g)\frac{\partial}{\partial g}
\ln{\cal X}^\sigma_{\rm S}(g)\,.
\ee
Denoting the coefficients in the perturbative expansion 
of $\gamma^\sigma_{\rm S}$ in the $\msbar$ 
coupling (as in Eq.~(\ref{gamma})) by $\gamma^\sigma_{\rm S;k}$, the
lowest order coefficients are given by 
\bea
\gamma^{\sigma}_{\rm S;0}&=&\gamma^{\sigma}_0\,,
\\
&&\nonumber
\\
\gamma^{\sigma}_{\rm S;1}&=&\gamma^{\sigma}_1-2b_0x^{\sigma}_{\rm S;1}\,.
\label{gammat1s}
\eea

For completeness we mention that for the lattice computations it is
often convenient to write the expressions in Sect.~\ref{sec_RGI} in
terms of lattice couplings. In this case, as an intermediate step, it
is useful to know the relation of the $\msbar$ coupling to the bare
lattice coupling: 
\be
{g}^2_{\msbar}(ta^{-1})=g_0^2+\frac{d_1(t)}{4\pi}g_0^4+\dots\,.
\label{gmstolatt}
\ee
To obtain the anomalous dimensions expressed in terms of the lattice
coupling to two-loop level, it is sufficient to know the coefficient
$d_1$ which takes the form:
\be
    d_1(t)=-8\pi b_0\ln(t)+d_{10}+d_{11}\Nf\,. \label{d1t}
\ee
The term $d_{10}$ depends on the pure gauge action;
e.g. for the Wilson gauge action it was computed
long ago by A.~and P.~Hasenfratz \cite{HaHa} and is given by
\bea
d_{10}&=&-\frac{\pi}{2\Nc}+k_1 \Nc\,,
\\
k_1&=&2.135730074078457(2)\,.
\eea
The term $d_{11}$ in (\ref{d1t}) for GW fermions depends on the parameter 
$s$ and is given by
\be
d_{11}(s)=4\pi 
\left[\frac{5}{72\pi^2}+d_{1,1}(\rho)+d_{1,2}(\rho)\right]\,,
\ee
where numerical values for the functions $d_{1,i}(\rho), i=1,2$,
with $\rho=1+s$, are given in Table~1 of ref.~\cite{APV}.
Corresponding values for $d_{11}(s)$ are given in Table~\ref{tabx}.

\begin{table}[ht]
\begin{center}
\begin{tabular}{|r|r|r|c|}\hline
\rule[-0.5ex]{0ex}{3.2ex}
$s\,\,\,$&$X_0\,\,\,\,\,\,$&$X_1\,\,\,\,\,\,$&$d_{11}$\\
\hline
$-0.8$&$-462.169$&$-472.573$&$0.287181$\\
$-0.7$&$-292.939$&$-302.192$&$0.256186$\\
$-0.6$&$-209.059$&$-217.347$&$0.235334$\\
$-0.5$&$-159.227$&$-166.669$&$0.219891$\\
$-0.4$&$-126.366$&$-133.048$&$0.207873$\\
$-0.3$&$-103.174$&$-109.157$&$0.198305$\\
$-0.2$&$ -86.004$&$ -91.337$&$0.190679$\\
$-0.1$&$ -72.836$&$ -77.557$&$0.184743$\\
$\phantom{-}0.0$&$-62.460$&$-66.599$&$0.180413$\\
$\phantom{-}0.1$&$-54.108$&$-57.689$&$0.177735$\\
$\phantom{-}0.2$&$-47.270$&$-50.313$&$0.176883$\\
$\phantom{-}0.3$&$-41.593$&$-44.113$&$0.178175$\\
$\phantom{-}0.4$&$-36.827$&$-38.836$&$0.182126$\\
$\phantom{-}0.5$&$-32.787$&$-34.295$&$0.189537$\\
$\phantom{-}0.6$&$-29.338$&$-30.351$&$0.201685$\\
$\phantom{-}0.7$&$-26.376$&$-26.896$&$0.220694$\\
$\phantom{-}0.8$&$-23.820$&$-23.848$&$0.250421$\\
\hline
\end{tabular}
\caption{\label{tabx}Values of functions $X_i$ appearing in 
\eq{xt1latt}, and $d_{11}(s)$ appearing in \eq{d1t}.}
\end{center}
\end{table}

%% file: appendix4.tex
\section{Correlators in Chiral Perturbation Theory \label{app4}}

In this appendix we describe the calculation of correlation functions
of left-handed charges and operators $\widehat{\cal O}_1^\sigma$ for a
finite periodic box with volume $V=T\cdot{L_1}\cdot{L_2}\cdot{L_3}$,
at order $\epsilon^2$ in ChPT. The same calculation in the
SU(3)-symmetric case has already been published in
ref.~\cite{HerLai02}, which can be consulted for further details. In
particular, we note that the sets of diagrams which appear at this
order in the $\epsilon$-expansion, are identical to the SU(4)-case
considered here, and are depicted in Figs.~1 and~3 of \cite{HerLai02}.

The result for the correlation function ${\cal C}^{ab}$ at order
$\epsilon^2$ is
\bea
  {\cal C}^{ab} &=& \Tr(T^a T^b)\frac{F^2}{2T}
   \bigg\{ 1+\frac{\Nf}{F^2}\bigg[\frac{\beta_1}{V^{1/2}}
  -\frac{T^2 k_{00}}{V}\bigg] \nonumber \\
  & & +\frac{2\Sigma T^2}{\Nf F^2}
  \left\langle
  \Re\Tr[MU_0\rme^{i\theta/\Nf}]\right\rangle_{\theta,U_0}
  h_1\Big(\frac{x_0}{T} \Big)\bigg\},
\eea
where $\Nf=4$, and the function $h_1$ is given by \cite{HasLeut90}
\be
   h_1(\tau)=\half\Big[\big(|\tau|-\half\big)^2
   -{\textstyle\frac{1}{12}}\Big].
\ee
The definitions of the coefficients $\beta_1$ and $k_{00}$, which
depend on the geometry of the box, can be found in
\cite{Hansen90}. Examples of numerical values are listed in
Table~\ref{tab_shape}.

\begin{table}
\begin{center}
\begin{tabular}{l r@{.}l l}
\hline
$T/L$ & \multicolumn{2}{c}{$\beta_1$} & $k_{00}$ \\
\hline
32/32 &    0&14046 & $0.07023=\beta_1/2$ \\
32/28 &    0&13872 &  0.07826 \\
32/24 &    0&13215 &  0.08186 \\
32/20 &    0&11689 &  0.08307 \\
32/16 &    0&08360 &  0.08331 \\
32/12 &    0&00582 &  0.08333 \\
32/8  & $-0$&21510 &  0.08333 \\
\hline
\end{tabular}
\caption{\footnotesize Some numerical values for $\beta_1$ and
$k_{00}$, for geometries with
$L_1=L_2=L_3\equiv{L}$. \label{tab_shape}}
\end{center}
\end{table}

For $[\widehat{\cal C}_1^\pm(x_0,y_0)]^{ab}_{\alpha\beta\gamma\delta}$
the general expression reads
\bea
[\widehat{\cal C}_1^{\pm}(x_0,y_0)]^{ab}_{\alpha\beta\gamma\delta}
    &=& \hat{\Delta}^{ab,\sigma}_{\alpha\beta\gamma\delta}
    \frac{F^4}{4T^2}\bigg\{1+(\Nf+\sigma)\frac{2}{F^2}
    \bigg[\frac{\beta_1}{V^{1/2}}-\frac{T^2 k_{00}}{V}\bigg]
    \nonumber \\
    & & +\frac{2\Sigma T^2}{\Nf F^2}\left\langle
        \Re\Tr[MU_0\rme^{i\theta/\Nf}] \right\rangle_{\theta,U_0}
        \bigg[h_1\Big(\frac{x_0}{T}\Big)+h_1\Big(\frac{y_0}{T}\Big)
        \bigg]\bigg\}.
\eea
Contact terms have been dropped in this expression, and we may assume
that $x_0,\,y_0$ are on opposite sides relative to the origin, i.e.
$x_0<0, y_0>0$. The flavour structure is encoded in the tensor
\bea
   \hat{\Delta}^{ab,\sigma}_{\alpha\beta\gamma\delta} &\equiv&
   \half\Big( T_{\gamma\alpha}^{\{a} T_{\delta\beta}^{b\}}
      +\sigma T_{\gamma\beta}^{\{a} T_{\delta\alpha}^{b\}}\Big)
   +\frac{1}{(4+\sigma)(4+2\sigma)}
   \Big( \delta_{\gamma\beta}\delta_{\delta\alpha}
  +\sigma\delta_{\gamma\alpha}\delta_{\delta\beta} \Big) \Tr(T^a T^b)
   \nonumber \\
& &-\frac{1}{2(4+2\sigma)}\Big( 
         \delta_{\gamma\beta}\{T^a,T^b\}_{\delta\alpha}
        +\delta_{\delta\alpha}\{T^a,T^b\}_{\gamma\beta} \nonumber\\
& &\phantom{-\frac{1}{2(4+2\sigma)}}
  +\sigma\delta_{\gamma\alpha}\{T^a,T^b\}_{\delta\beta}
  +\sigma\delta_{\delta\beta}\{T^a,T^b\}_{\gamma\alpha} \Big),
\eea
where
\be
   T_{\gamma\alpha}^{\{a} T_{\delta\beta}^{b\}} \equiv
   T^a_{\gamma\alpha} T^b_{\delta\beta} +
   T^a_{\delta\beta} T^b_{\gamma\alpha}.
\ee
In order to study the case of a light charm quark, one can choose a
diagonal, degenerate mass matrix and generators such that
\be
   T^a_{\tau\omega}=\delta_{\tau\gamma}\delta_{\omega\alpha},\qquad
   T^b_{\tau\omega}=\delta_{\tau\delta}\delta_{\omega\beta},
\label{eq_TaTb}
\ee
where $\alpha,\beta,\gamma,\delta$ are assumed all fixed and
different. In this case one finds that $\hat{\Delta}^{ab}=1/2$, and
for a symmetric spatial volume and vanishing vacuum angle $\theta$ one
recovers the expressions in eqs.~(\ref{eq_Cab_res})
and~(\ref{eq_C1_res}).

Correlation functions in sectors of fixed topology, characterised by
an index $\nu$, can be obtained from those for fixed vacuum angle
$\theta$ via Fourier transform. If we assume a diagonal mass matrix
and define
\be
   Z_\theta(u/2) = \int_{U_0}\rme^{(u/2)\Re\Tr[U_0\exp(i\theta/\Nf)]},
   \quad
   Z_\nu(u/2) = \frac{1}{2\pi}\int_0^{2\pi}\rmd\theta\,
   \rme^{-i\nu\theta}Z_\theta(u/2),
\ee
the quantity $C_\theta(u/2)$ in eqs.~(\ref{eq_Cab_res})
and~(\ref{eq_C1_res}) gets simply replaced by
\be
   C_\theta(u/2)\equiv\frac{2}{\Nf}\frac{\partial}{\partial u} \ln
   Z_\theta(u/2) 
   \longrightarrow
   \frac{2}{\Nf}\frac{\partial}{\partial u} \ln Z_\nu(u/2) 
   \equiv C_\nu(u/2).
\label{eq_Ctheta_Cnu}
\ee
In the small and large mass regions, $C_\nu(u/2)$ is approximately
given by
\be
   C_\nu(u/2)\approx\left\{
   \begin{array}{cl}
       2|\nu|/u, & u\ll1, \\
       1,        & u\gg1.
   \end{array}\right.
\ee